\documentclass[10pt,preprint]{aastex}
\newcommand{\ee}[1]{\mbox{${} \times 10^{#1}$}}

\begin{document}

\title{Anisotropic Galactic Outflows and Enrichment of the Intergalactic 
Medium. I: Monte Carlo Simulations}

\author{Matthew M. Pieri\altaffilmark{1}, Hugo Martel\altaffilmark{1},
and C\'edric Grenon\altaffilmark{1,2}}

\altaffiltext{1}{D\'epartement de physique, g\'enie physique et optique,
Universit\'e Laval, Qu\'ebec, QC, G1K 7P4, Canada}

\altaffiltext{2}{Hubert Reeves fellow}

\begin{abstract}

We have developed an analytical model to describe the evolution
of anisotropic galactic outflows. With it, we investigate the 
impact of varying
opening angle on galaxy formation and the evolution of the intergalactic
medium. We have implemented this model in
a Monte Carlo algorithm to simulate galaxy formation and outflows in
a cosmological context. Using this algorithm, we have simulated the evolution
of a comoving volume of size $(12h^{-1}{\rm Mpc})^3$ in the $\Lambda$CDM
universe. Starting from a Gaussian density field at redshift $z=24$,
we follow the formation of $\sim20,000$ galaxies and simulate the galactic
outflows produced by these galaxies. When these outflows collide with
density peaks, ram pressure stripping of the gas inside the peaks may result. 
This occurs in around half the cases and prevents the formation of galaxies. 
Anisotropic outflows follow the path of least resistance and thus
travel preferentially into low-density regions, away from cosmological
structures (filaments and pancakes) where galaxies form. As a result,
the number of collisions is reduced, leading to the formation of a larger
number of galaxies. Anisotropic outflows can significantly enrich low-density
systems with metals. Conversely, the cross-pollution in metals of objects
located in a common cosmological structure, like a filament, is
significantly reduced. Highly anisotropic outflows can 
travel across cosmological voids and deposit metals in other,
unrelated cosmological structures.

\end{abstract}


\section{INTRODUCTION}

Galactic outflows play an important role in the evolution of galaxies
and the intergalactic medium (IGM). Supernova explosions in galaxies
create galactic winds, which deposit energy and metal-enriched gas into the
IGM. These outflows are necessary to explain many observations and to
solve many problems in galaxy formation, such as
the high mass-to-light ratio of dwarf galaxies, the observed metallicity 
of the IGM, the entropy content of the IGM, the abundance of dwarf galaxies
in the Local Group, the overcooling problem,
and the angular momentum problem.

High-resolution, gasdynamical simulations of explosions in a single
object reveal that outflows generated by such explosions tend
to be highly anisotropic, with the energy and metal-enriched
gas being channeled along the direction of least resistance, where
the pressure is the lowest  \citep{mf99,ms01a,ms01b}. Furthermore,
several observations support the existence of anisotropic
outflows 
(e.g. \citealt{bt88,fhk90,carignanetal98,sbh98,stricklandetal00,vr02}).

Indirect support for the existence of anisotropic outflows comes from
the observed enrichment of systems around the mean density of the 
universe \citep{2003ApJ...596..768S,2004MNRAS.347..985P}
and the enrichment of systems far from known galaxies at $z\sim3$ 
\citep{psa06, s06}. 
It may be challenging to enrich such regions with
isotropic outflows even with the inclusion of enrichment from poorly 
understood Population III stars. Early indications are that 
Population III stars
 are unlikely 
to pollute the IGM to a large extent \citep{nop04}. Anisotropic
outflows may also provide an explanation for part of 
the observed scatter in the metallicity in the IGM, which is still unexplained
 \citep{2003ApJ...596..768S,psa06}.

Several simulations of galactic outflows in cosmological contexts
have been performed. Typically, these simulations use cubic comoving
volumes of size $\sim(10\,{\rm Mpc})^3$, containing thousands of
galaxies. The methods used can be divided into two groups
: analytical 
methods and numerical methods. 
The analytical methods describe the expansion of outflows in simulations
that are either Gaussian random realizations of the density power spectrum
or N-body simulations combined with prescriptions for galaxy formation.
Outflows are represented using an analytical
solution (e.g. \citealt{sb01}, hereafter SB01, \citealt{tms01,bsw05}) 
that currently assume isotropy. 

The numerical simulations 
use hydrodynamical algorithms such as SPH, 
and outflows are generated in a variety of ways: by imparting SPH particles with a large 
velocity component (e.g. \citealt{std01,sh03,od06}), depositing
additional thermal energy into SPH particles 
(e.g. \citealt{2002ApJ...578L...5T}), or taking output from
completed SPH simulations and calculating, {\it a posteriori},
the propagation of outflows into the IGM \citep{aguirreetal01}. Unlike
the analytical methods cited above, which assume isotropic outflows, all
these numerical approaches have the potential to generate anisotropic
outflows. However, we believe that there are some limitations to these
numerical approaches, which motivates us to introduce an analytical model
for anisotropic outflows.

With any SPH simulation, we need to be concerned with the limited resolution
of the algorithm. Consider for example the simulations of \citet{std01}.
These authors identify galaxies of radius $r_N$, and rearrange
the SPH particles located between $r_N$ and $r_0=2r_N$ into
two uniform, concentric spherical shells located at radii $0.9r_0$ and $r_0$, 
which are
then given an outward velocity. The outflow is therefore initially isotropic
but becomes anisotropic as it propagates into a non-uniform external medium. 
We can see two potential problems with this
approach. Firstly, the structure responsible for generating the anisotropy
might be the galaxy or its environment, in which case the rearranging of particles
into concentric shells would erase that structure entirely. Secondly, 
52 particles per shell (the number they used) provides a good covering of the
solid angles initially, but as the outflow expands and the particles in the
shells move apart they become more like individual pressure points pushing
on the external medium, and this could lead to an artificial 
mixing of the outflow and the external medium by Raleigh-Taylor
instability.

The approach of \citet{aguirreetal01} is radically different. It consists
of identifying galaxies in an output from an SPH simulation and calculating
the propagation of outflows from these galaxies in $N_a$ different directions.
Since the resistance encountered by the outflows will be direction-dependent,
outflows will start isotropic but then become anisotropic as the distance
travelled by outflows will vary with direction. There are two limitations
to this approach. Firstly, since it uses the output of an SPH simulation and
introduces the outflows {\it a posteriori}, the feedback effect of these 
outflows cannot be simulated (i.e. outflows do
not influence the formation of other galaxies). Secondly,
in this approach gas elements move radially and encountering high-pressure
gas will dissipate their energy and rapidly slow down. In the real universe,
that gas element will likely acquire a tangential velocity component that will redirect
it towards regions where the resistance of the external medium is 
weaker.

To overcome these various limitations and study anisotropic outflows in
a cosmological context, {\it including the effect of feedback},
we have designed an analytical model for galactic outflows.
This can then be combined with either an analytical method
or a semi-analytic method for the description of galaxy formation. In this
paper, we use an analytical Monte Carlo method. Calculations performed
with an N-body semi-analytic approach will be presented in a forthcoming 
paper \citep{mgp07}.

This paper is set out as follows. In \S2, we describe our analytical model
for anisotropic outflows. In \S3, we describe our Monte Carlo method
for cosmological simulations. Results are presented in \S4, and their
implications are discussed in \S5. Conclusions are presented in \S6.

\section{A MODEL FOR ANISOTROPIC OUTFLOWS}

Consider a density field with a local density maximum at some position,
$P$. A halo may collapse at that point and may go on to form a galaxy, which will then
produce an outflow. Our goal is to design an analytical model for the
geometry and evolution of this outflow that takes into account the physical
properties of the galaxy, the density distribution of the matter
surrounding the galaxy, and the global properties of the IGM.

\begin{figure}[th]
\hskip0.8in
\includegraphics[width=5.in]{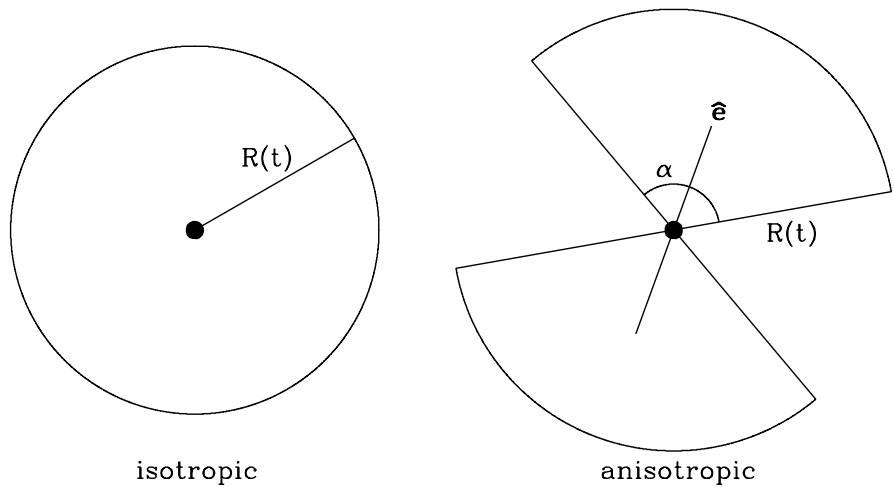}
\caption{Geometry of the isotropic and anisotropic outflows. The isotropic
outflows are spherical; the anisotropic ones are bipolar spherical cones.
$R(t)$ is the time-dependent radius of the outflow, $\alpha$ is the opening
angle, and $\hat{\bf e}$ the direction of the outflow.}
\label{schematic}
\end{figure}

Using the high-resolution simulations and the observations of galactic
outflows as a guide, we will  
represent outflows as ``bipolar spherical cones.'' In a 
spherical coordinate system $(r,\theta,\phi)$, the outflow occupies
the volume defined by $r\leq R(t)$, $\theta\leq\alpha/2$ or 
$\theta\geq\pi-\alpha/2$, and $0\leq\phi<2\pi$
where $R(t)$ is the radius of the outflow
and $\alpha$ is the opening angle. This is illustrated in 
Figure~\ref{schematic}.

In the limit $\alpha=\pi$, the outflow becomes
spherical, and is entirely described by the radius, $R(t)$. If $\alpha<\pi$,
the outflow is anisotropic, and two additional parameters must be specified:
the opening angle $\alpha$, and the direction of the outflow, which is
defined by a unit vector $\hat{\bf e}$. Hence, our
model of anisotropic outflows is a 3-parameter model.

One possible way to describe our model phenomenologically is to imagine
that the outflow starts isotropic, with all the matter moving radially, but
the parts of the outflow that encounter high-resistance from the external 
medium acquire a transverse velocity and get redirected into the directions
where the resistance is weaker. In this context, we see that our model and
the model of \citet{aguirreetal01} represent two limiting cases. In their
model, the energy of the gas expanding into dense regions is entirely
dissipated. In our model, that energy is entirely funneled into the
less-dense regions.

\subsection{The Opening Angle}

It is not clear what value one should be using for the opening angle.
The simulations of \citet{mf99} and \citet{ms01a} show that the angle actually
varies as the expansion proceeds. 
It starts at a low value, $\alpha\sim10^\circ-45^\circ$
near the site of the explosion. Once the outflow reaches the low-density
(around or below the mean density of the Universe) regions it ``fans out'', and the opening angle increases to values 
$\alpha\sim45^\circ-100^\circ$, or even larger. The radius also varies
with opening angle. 
Hence, our representation of outflows as bipolar spherical
cones of fixed opening angle is a convenient simplification. 
Until we have a better  understanding of the morphology of the
outflows (something that would require more precise simulations), we will
treat the opening angle as a free parameter that can take any value from
$\alpha=180^\circ$ (the isotropic limit) to $\alpha\sim20^\circ$
(the jets limit), keeping in mind that the larger values are likely to
provide a more accurate description of the outflows once they reach the
low-density regions. 

It is reasonable to expect that the opening angle
will vary for individual outflows on a case-by-case basis due to
their differing driving pressure. We treat the chosen value as a typical 
opening angle with an aim to reproducing the global properties 
of anisotropic outflows for a cosmological volume.

\subsection{The Direction of the Outflow}
\label{direction outflow}

The outflows are expected to take a path of least resistance out of a
galaxy that forms at the location of a 
density peak. In the current literature, two 
characteristic scales for this path of least resistance have been
considered: the galactic scale, in which a disk may form \citep{mf99}, 
and larger scale filamentary or pancake structures \citep{ms01a,ms01b}. 
In the former case, outflows are directed along the rotation axis of the
galaxy. In the latter the pressure of the surrounding medium determines
the direction of the outflow. We have
selected the latter as the defining scale, upon which our anisotropic 
outflows are based, for the following reasons:

\begin{itemize}

\item{It is unclear if the galaxy has already
formed a well-ordered disk by the time 
much of the initial starburst has occurred. Outflows may start while the
galaxy is still in the process of assembling. Hydrodynamical simulations
of galaxy formations suggest that starbursts take place during
a chaotic merger events at high redshift, before a disk has
assembled \citep{brooketal05,brooketal06}.
}

\item{Despite recent results suggesting their rotation
axes tend to lie preferentially parallel to the plane of 
pancake structures 
(\citealt{nl06}, and references therein) the orientation of 
these disks are largely
randomized. Averaging over all outflowing galaxies, only a 
weak directionality is 
expected before the effect of larger scale structures begins to dominate as the
outflows continue to expand. }

\item{ The importance of disk-scale effects is dependent on the 
locations of SNe within
the disk. Bipolar outflows due to SNe explosions in a disk are a result of the 
relatively high pressure gradient along the rotational (minor) axis compared
to the major axis. It seems reasonable to assume, therefore, that for 
off-center SNe, where the relative difference in pressure gradient is less 
acute, the outflow will be less bipolar.}

\end{itemize}

\begin{figure}[t]
\hskip0.25in
\includegraphics[width=6.0in]{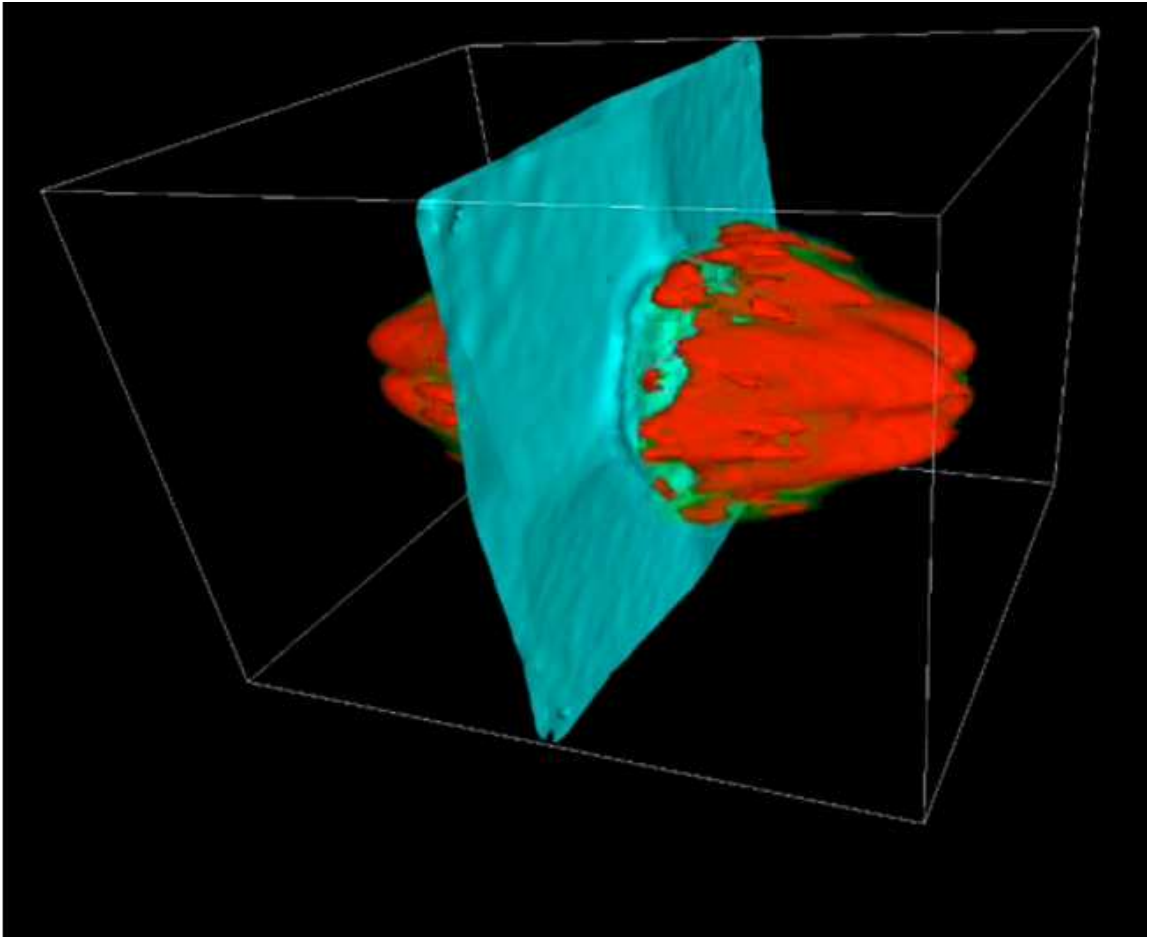}
\caption{Snapshot of a hydrodynamical simulation of an explosion inside
a cosmological pancake. Blue: density isosurface, showing the pancake.
Notice the circular ripple resulting from the explosion.
Red: temperature isosurface, showing the outflow.}
\label{pancake}
\end{figure}

Figure~\ref{pancake} shows an intermediate stage of a simulation
\citep{ms01a}
of an explosion inside a dwarf galaxy that is forming at the intersection
of two emerging filaments inside a cosmological pancake. This simulation
was performed with an Adaptive SPH algorithm \citep{smvo96,ovsm98} with
$64^3$ gas and $64^3$ dark-matter particles. The outflow is clearly
anisotropic, bipolar, and propagates in the direction normal to the pancake 
plane. Interestingly, the central region where the explosion takes place
has a nearly isotropic density profile. Hence, it is the anisotropy of
the outer regions that results in an anisotropic outflow.

In our Monte Carlo simulations,
we assume that galaxies form at the location of local density peaks in
the matter distribution, and we determine the direction of the outflow
by finding the direction of least resistance in the vicinity of that peak. 
We discuss in greater detail the concepts
of ``peaks'' and ``vicinity'' in \S\S3.2 and 3.3 below.

Consider a local density maximum located at some point, $P$, inside
the computational volume. We center a cartesian coordinate system
$(x,y,z)$ on that point, and perform a Taylor expansion of the density
contrast up to second order,
\begin{equation}
\label{del1}
\delta(x,y,z)=\delta_{\rm peak}-Ax^2-By^2-Cz^2-2Dxy-2Exz-2Fyz\,.
\end{equation}

\noindent This expression contains no linear terms since the density
is a local maximum. In practice, we consider the density distribution
inside a sphere of radius, $R^*$, centered on
the point, $P$, and perform a least-square fit of 
equation~(\ref{del1}) to determine numerically the values of the 6 
coefficients $A$, $B$, $C$, $D$, $E$, $F$.
Once these coefficients are determined, we rotate to a new 
coordinate system $(x',y',z')$, such that the 
the cross-terms vanish. Equation~(\ref{del1}) reduces to
\begin{equation}
\delta(x',y',z')=\delta_{\rm peak}-A'{x'}^2-B'{y'}^2-C'{z'}^2\,.
\end{equation}

\noindent It is actually straightforward to perform this
change of coordinates. We combine the coefficients of equation~(\ref{del1}) to
form the following matrix,
\begin{equation}
M=\left[\matrix{A&D&E\cr D&B&F\cr E&F&C\cr}\right]\,.
\end{equation}

\noindent We then diagonalize this matrix. Since this matrix is real and
symmetric, all three eigenvalues are real. These eigenvalues are the
coefficients $A'$, $B'$, $C'$, respectively,
and the 3 corresponding eigenvectors give us the directions of the three
coordinate axes $x'$, $y'$, and $z'$, respectively.

The three coefficients $A'$, $B'$, and $C'$ are always positive,
otherwise the point, $P$, would not be a maximum, but rather a saddle point
or a minimum. Each coefficient is a measure of how fast the density
drops as we move away from the peak in the direction corresponding
to that coefficient. Therefore, the largest coefficient corresponds to
the direction along which the density drops the fastest. We will interpret
this as being the direction of least resistance, and assume that the outflow
will naturally follow that direction. For instance, if $B'$ happens to be
the largest of the 3 coefficients, then the outflow will be directed along the
$y'$-axis.

\subsection{The Expansion of the Outflow}
\label{expansion of outflow}

In this section, we present the equations describing the expansion of the
outflow. The technique used for solving these equations is shown in
Appendix~A.

\subsubsection{Basic Equations}

Tegmark, Silk, \& Evrard (1993, hereafter TSE) presented
a formulation of the expansion of isotropic outflows that
was based on previous work by numerous authors 
\citep{cs74,mko77,weaveretal77,mcs79,bruhweileretal80,thi80,mck87,omk88} 
and refined it to account for the expansion of the universe.
In this formulation, the injection of thermal energy
produces an outflow of radius, $R$, which consists of a dense shell
of thickness $R\delta$ containing a cavity.
A fraction, $1-f_m$, of the gas is piled up in the shell, while a
fraction, $f_m$, of the gas is distributed inside the cavity.
We normally assume $\delta\ll1$, $f_m\ll1$, that is, 
most of the gas is located inside a thin shell.
This is called the {\it thin-shell approximation}.

The original equations of TSE have been refined by several authors
(\citealt{mfr01}; SB01; \citealt{sfm02}, hereafter SFM) 
to include various additional physical
processes. Hence, there is not a ``unique'' form of the equations.
In this paper, the evolution of the shell radius, $R$, expanding out of a halo
of mass, $M$, is described
by the following system of equations,
\begin{eqnarray}
\label{rdotdot}
\ddot R&=&{8\pi G(p-p_{\rm ext})\over\Omega_bH^2R}-{3\over R}(\dot R-HR)^2
-{\Omega H^2R\over2}-{GM\over R^2}\,,\\
\label{pdot}
\dot p&=&{L\over2\pi R^3}-{5\dot Rp\over R}\,,
\end{eqnarray}

\noindent where a dot represents a time derivative, $\Omega$, $\Omega_b$,
and $H$ are the total density parameter, baryon density parameter, and
Hubble parameter at time, $t$, respectively,
$L$ is the luminosity (discussed in \S 2.3.3), $p$ is the bubble pressure
resulting from this luminosity, and $p_{\rm ext}$ is the external
pressure of the IGM. 
The four terms in equation~(\ref{rdotdot}) represent, from left to right, the 
driving pressure of the outflow, drag due to sweeping up the IGM and 
accelerating it from velocity $HR$ to velocity $\dot R$, and the gravitational
deceleration caused by both the expanding shell and the halo itself.
The two terms in equation~(\ref{pdot}) represent the increase
in pressure caused by injection of thermal energy, and the drop in pressure caused by
the expansion of the outflow, respectively. For details, we refer the reader
to \citet{omk88} and TSE. 

These expressions most closely resemble that of 
SFM, the one difference coming from their use of a radially 
dependent mass 
in the last term of equation~(\ref{rdotdot}),
derived by assuming a NFW profile \citep{nfw97}. This is
motivated by the gravitational drag on a spherical outflow expanding out of a 
spherical collapsed halo,
and is no longer an adequate description in the case being examined since 
an anisotropic outflow will more rapidly escape an elliptical halo. The 
influence 
of the gravitational potential well will be an intermediate case between 
this and
a point mass. In practice we take the simpler form by assuming a point mass,
as in SB01. 
We justify this approximation as follows: early in the outflow lifetime the effect of
gravity on the outflow is negligible compared to other effects 
(see Appendix~A). As the outflow expands, the gravity terms eventually become
important, by which time most outflows have left the halo from which
they originate.
Notice that the original formulation of TSE did not include
the term $-GM/R^2$, and also ignored the external pressure $p_{\rm ext}$.

These equations provide a full description of isotropic outflows. We now
modify these equations to describe anisotropic outflows
 with any value of the opening angle, $\alpha$. We impose the
condition that the modified equations reduce to equations~(\ref{rdotdot})
and~(\ref{pdot}) in the isotropic limit $\alpha=180^\circ$.

First consider
equation~(\ref{pdot}). The two terms in the right hand side correspond
to the increase in pressure caused by thermal energy injection, and the decrease in
pressure caused by the expansion of the outflow, respectively. At this point,
we need to backtrack. Equation~(\ref{pdot}) was derived from the following
system of equations (TSE):
\begin{eqnarray}
\label{et}
E_t&=&{3pV\over2}\,,\\
\label{etdot}
\dot E_t&=&L-p\dot V\,,
\end{eqnarray}

\noindent where $E_t$ is the thermal energy inside the outflow,
and $V$ is the volume of the outflow. We take the time derivative
of equation~(\ref{et}), and eliminate $\dot E_t$ using equation~(\ref{etdot}).
We get
\begin{equation}
\label{pdotgen}
\dot p={2L\over3V}-{5p\dot V\over3V}\,.
\end{equation}

\noindent If we now substitute $V=4\pi R^3/3$, we recover 
equation~(\ref{pdot}). However, a bipolar anisotropic outflow with opening 
angle, $\alpha$, has a volume given by
\begin{equation}
\label{volume}
V={4\pi R^3\over3}\left(1-\cos{\alpha\over2}\right)\,.
\end{equation}

\noindent We substitute equation~(\ref{volume}) in equation~(\ref{pdotgen}),
and get
\begin{equation}
\label{pdotnew}
\dot p={L\over2\pi R^3[1-\cos(\alpha/2)]}-{5\dot Rp\over R}\,.
\end{equation}

\noindent Physically, as $\alpha$ decreases, the energy is injected in a 
smaller volume, resulting in a larger pressure, $p$. This equation replaces
equation~(\ref{pdot}). In our model, this is the only
modification we need to make to the equations describing the
expansion of the outflow (although a modification to the luminosity 
will be required). 
Equation~(\ref{rdotdot}) remains unchanged.

\subsubsection{The External Pressure}

The external pressure, $p_{\rm ext}$, depends upon the density and
temperature of the IGM, which can be quite complex. But in the context
of the analytic approximation we are using, we are justified
in making some simplifying approximations. Following \citet{mfr01}, we will
assume a photoheated IGM with a fixed temperature, 
$T_{\rm IGM}=10^{4}{\rm K}$. We will also
assume an IGM density, $\rho_{\rm IGM}$, equal to the mean baryon density,
$\bar\rho_b$.
\citet{hg97} performed detailed
hydrodynamical calculations of the equation of state for a photoheated
IGM, and showed that the temperature is a function of the density.
But according to their Figure~1, a temperature of $10^4{\rm K}$ is
appropriate for densities, $\rho\approx\bar\rho_b$. 
It should also be noted that the temperature of the IGM is redshift-dependent
as shown by \citet{s00} and we will neglect this here. Our choice of fixed 
external temperature and density will be discussed further in \S 5.
The external pressure
is then given by
\begin{equation}
\label{pext}
p_{\rm ext}(z)={\bar\rho_bkT_{\rm IGM}\over\mu}=
{3\Omega_{b,0}H_0^2kT_{\rm IGM}(1+z)^3\over8\pi G\mu}\,,
\end{equation}

\noindent where $z$ is the redshift, $\mu$ is the mean molecular mass,
and subscripts 0 indicate present values. The value of $\mu$
depends upon the ionization state of the gas.  We assume that hydrogen
is ionized \citep{p06} and helium is singly-ionized in the redshift range 
$3<z<7$ \citep{t02,cf05}, which is most of the redshift range in which our
 outflows are actively expanding. Also, even when much of the IGM is still neutral,
 it is reasonable to expect the source galaxy to have photoionized the region
ahead of the outflow. We will assume that this is the case 
throughout the simulations. The mean molecular weight is then
$\mu=2/(4-3Y)\,{\rm a.m.u.}$ (atomic mass units).
For a helium abundance, $Y=0.242$ \citep{it04}, we get 
$\mu=0.611\,{\rm a.m.u.}$.

\subsubsection{The Luminosity}

The luminosity, $L$, is the rate of energy deposition or 
dissipation within the 
outflow and is a sum of five terms,
\begin{equation}
L(t)=L_{\rm SN}-L_{\rm comp}-L_{\rm ne^2}-L_{\rm ion}+L_{\rm diss}\,,
\end{equation}

\noindent where $L_{\rm SN}$ is the total luminosity of the supernovae 
responsible for generating  the outflow, $L_{\rm comp}$ represents
the cooling due to Compton drag against CMB photons, $L_{\rm ne^2}$ 
represents the cooling due to 2-body interactions, $L_{\rm ion}$ 
represents the cooling due to ionization, and $L_{\rm diss}$ 
represents heating from collisions between the shell and the IGM.
Following SFM, we will assume that the first two terms,
supernovae and Compton drag, dominate, and will neglect the remaining
terms. The supernovae luminosity, for a galaxy forming in a halo 
of mass, $M$, is given by
\begin{equation}
L_{\rm SN}={f_{\rm w}E_0 \over t_{\rm burst}} {M_*\over M_{\rm req}}
=2.86f_{\rm w}f_*\left({\Omega_{b,0}\over\Omega_0}\right) 
\left({M\over1{\rm M}_\odot}\right)\,{\rm L}_\odot\,,
\end{equation}

\noindent where $f_{\rm w}$  is the fraction of the total energy
released that goes into the outflow, $f_*$ is the star formation
efficiency, and 
the total mass in stars formed during the starburst is 
$M_*=f_*M\Omega_{b,0}/\Omega_0$. $M_{\rm req}$ is the mass of stars
required to form one SN and we take a value of 
$M_{\rm req}=89.7 {\rm M_\odot}$ (derived using a broken power-law IMF
from \citealt{k01}) and the energy released by each of these
SNe is 
$E_0=10^{51} {\rm ergs}$. \citet{lh95} show that for
instantaneous star formation, as considered here, the SN outflow phase 
associated
with this burst last for a duration of 
$t_{\rm burst}= 5\times10^{7}{\rm yr}$ and the delay
before the first SNe is negligible. We use the galaxy mass dependent 
determination of $f_{\rm w}$ given by SFM, 
$f_{\rm w}(M)=0.3\delta_B(M)/\delta_B(M = 2 \times 10^8 {\rm M_\odot})$,
where
\begin{equation}
\delta_B(M)= 
\cases{
1.0, & $\tilde{N}_t\le 1$;\cr
1.0+0.165\ln(\tilde{N}_t^{-1}), & $1 \le \tilde{N}_t \le 100$;\cr
\left[1.0-0.165\ln(100)\right]100\tilde{N}_t^{-1}, &  $100 \le \tilde{N}_t$;\cr
}
\end{equation}

\noindent
and $\tilde{N}_t\equiv 1.7 \times 10^{-7}
({\Omega_{b,0}/\Omega_0})M/(1{\rm M_\odot})$. 

The ionization state inside the outflow is even more uncertain and may
 be a function of distance from the source as the gas may recombine as it expands from
 its source galaxy. Until we have a
more precise model for the internal structure of the outflow, we will simply
make the same assumption as for the surrounding IGM: ionized hydrogen
and singly-ionized helium. However, it is worth noting that the extreme case, where
outflows contain fully ionized gas, involves only a minor modification of the Compton 
drag. Following TSE, the Compton luminosity is given by
\begin{equation}
\label{lcomp}
L_{\rm comp}={\pi^2\over15}(\sigma_tcn_e)\left({kT_e\over m_ec^2}\right)
(kT_\gamma)^4{V\over(\hbar c)^3}\,,
\end{equation}

\noindent 
where $\sigma_t$
is the Thompson cross section, $T_\gamma$ is the CMB temperature,
and $n_e$ and
$T_e$ are the electron number density and temperature inside the bubble, 
respectively. The pressure, $p$, inside the bubble is given by
\begin{equation}
\label{et1}
p=nkT_e=(n_e+n_{\rm ion})kT_eV\,.
\end{equation}

\noindent We assume that hydrogen is ionized and helium is 
singly-ionized and hence $n_{\rm ion}=n_e$, which is also the approximation 
made in TSE. Equation~(\ref{et1}) becomes $n_ekT_e=p/2$.
We now eliminate $n_ekT_e$
in equation~(\ref{lcomp}). 
We also eliminate $V$ using equation~(\ref{volume}), and get
\begin{equation}
\label{lcomp2}
L_{\rm comp}={2\pi^3\over45}{\sigma_t\hbar\over m_e}
\left({kT_{\gamma0}\over\hbar c}\right)^4
\left(1-\cos{\alpha\over2}\right)(1+z)^4pR^3\,,
\end{equation}

\noindent where
$T_{\gamma0}$ is the present CMB temperature. We used 
$T_{\gamma}=T_{\gamma0}(1+z)$, which is valid over the range
of redshifts we consider.

\subsubsection{The Final Stage of the Outflow}

The expansion of the outflow is initially driven by the supernovae luminosity.
After the supernovae turn off, the outflow enters the ``post-supernova phase.''
The pressure inside the outflow keeps driving the expansion, but this
pressure drops since there is no energy input from supernovae.
Eventually, the pressure will drop down to the level of the external IGM 
pressure. At that point, the expansion of the outflow will simply follow 
Hubble expansion. Since we assume identical mean molecular weights inside
and outside the outflow, the condition $p=p_{\rm ext}$ implies 
$T=T_{\rm IGM}=10^4{\rm K}$. 

\begin{figure}[t]
\hskip0.5in
\includegraphics[width=5.5in]{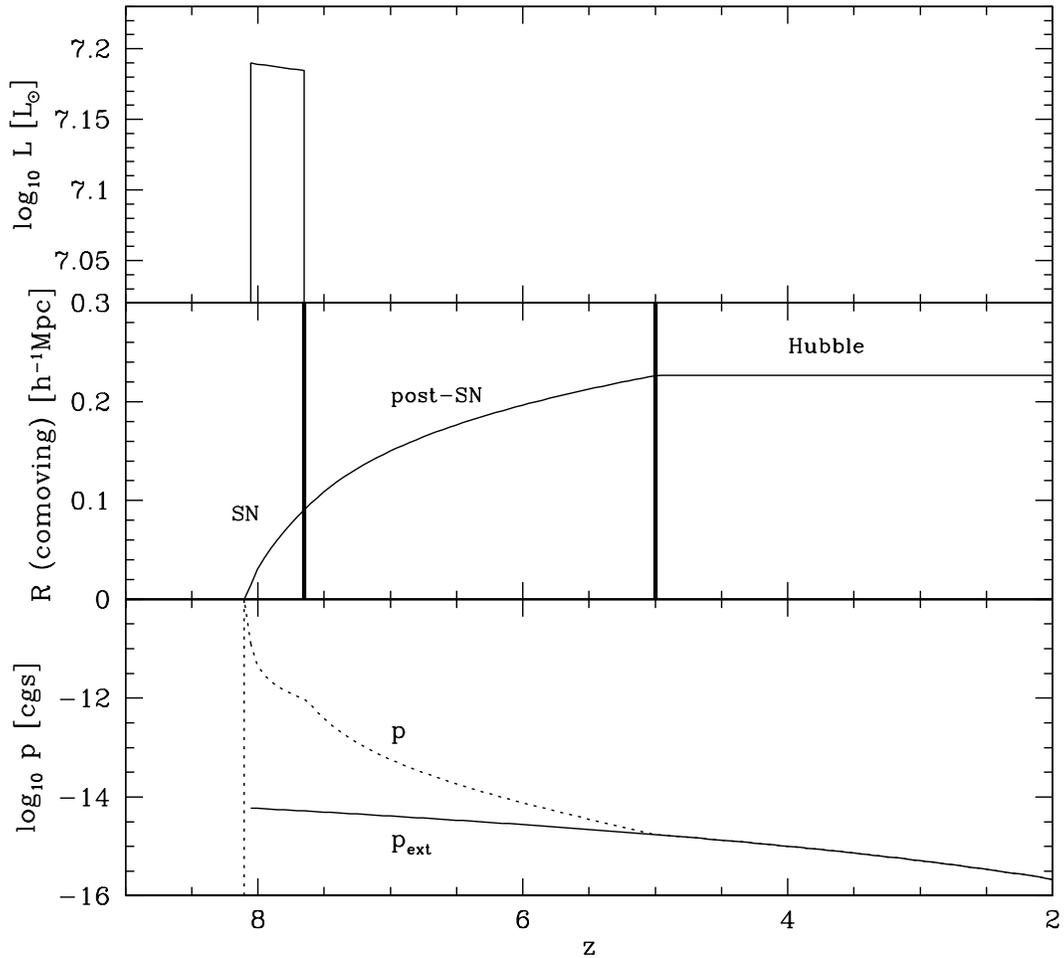}
\caption{Evolution of the largest outflow. The middle panel shows the
comoving radius of the outflow versus redshift, $z$. The thick lines
separate the various phases of the expansion (supernova-driven, 
post-supernova, 
Hubble expansion). The top panel shows the  total
luminosity, $L=L_{\rm SN}-L_{\rm comp}$, versus redshift. 
The bottom panel shows the external IGM pressure (solid line) and
the pressure of the outflow (dotted line).}
\label{out001}
\end{figure}

Figure~\ref{out001} shows the isotropic outflow solution for a halo of 
mass, $M=2.8\times10^9{\rm M}_\odot$ that
 collapses at redshift, $z=8.10$, in a $\Lambda$CDM universe. 
 The cooling time is short, $3 {\rm Myr}$, so the galaxy forms soon after,
 at $z=8.07$. The outflow turns on immediately, as discussed in the previous 
 section, and goes on to form the largest outflow in our simulations.
 
The top panel shows the total luminosity, and the middle
panel shows the comoving radius. The bottom panel shows the
pressure, $p$, of the outflow and the external pressure, $p_{\rm ext}$, of the
IGM. Initially, during the supernova phase, the outflow is driven by the
energy input from supernovae and the radius steadily increases. 
During this phase, the supernova luminosity,
$L_{\rm SN}$, exceeds the Compton luminosity by factors of several and 
 for this particular outflow dominates it entirely. The pressure
diverges at the redshift of
supernova turn-on (because we neglect the finite volume of the region
containing the supernovae), but then drops as $t^{-4/5}$ (see Appendix~A).
At $z=7.63$, the supernovae burn out, and the only contribution
to the total luminosity is the energy loss by Compton drag. The outflow
enters the post-supernova phase. The turn-off of the supernovae produces 
a change 
of slope in the pressure, which in turn produces a change of slope in
the Compton luminosity. The post-supernova phase lasts until redshift
$z=4.95$. At that point, the pressure of the outflow becomes equal to 
the external pressure, and the outflow switches to pure Hubble expansion.

\section{THE MONTE CARLO METHOD}

We use essentially the same Monte Carlo method as in SB01, generalized
to include anisotropic outflows. In our dynamical equation for
the evolution of the outflow [eq.~(\ref{rdotdot})], 
we include the external pressure, $p_{\rm ext}$, an effect that was neglected
in SB01, but included in SFM.

\subsection{Cosmological Model and Initial Conditions}

We consider a $\Lambda$CDM model with present
density parameter, $\Omega_0=0.27$, 
baryon density parameter, $\Omega_{b,0}=0.04444$, cosmological
constant, $\lambda_0=0.73$, Hubble constant,
$H_0=71\rm\,km\,s^{-1}Mpc^{-1}$ ($h=0.71$), 
primordial tilt,
$n_s=0.93$, and CMB temperature, $T_{\rm CMB}=2.725$, consistent with the
results of {\sl WMAP} \citep{bennetetal03}. 
We simulate structure formation inside a comoving cubic volume of 
size, $L_{\rm box}=12h^{-1}{\rm Mpc}$, with periodic boundary conditions.
To generate initial conditions, we lay down a cubic grid of size 
$512\times512\times512$ in the computational volume, and compute
the initial density contrast, $\delta_i$,
at initial redshift, $z_i=24$. 
We also compute, on 10 similar grids, the same
density contrast, but filtered at the 10 mass 
scales: $M_1,M_2,\ldots,M_{10}$. As we use a Gaussian filter,
the mass, $M$, and comoving radius, $R_f$, of the filter are
related by $M=(2\pi)^{3/2}R_f^3\bar\rho_0$, where
$\bar\rho_0=3\Omega_0H_0^2/8\pi G$ is the present mean density of
the universe.
The values of the mass and length filtering scales are listed in
Table~\ref{filtertab}. The last two columns indicate the filtering 
radius in units of the grid spacing, $\Delta$, and in units of
the box size, $L_{\rm box}$.

\begin{deluxetable}{llrrl}
\tablecaption{Mass and Length Filtering Scales}
\tablewidth{0pt}
\tablehead{
\colhead{Filter name} & \colhead{$M\;[{\rm M}_\odot]$} & 
\colhead{$R_f\;[{\rm kpc}]$} & \colhead{$R_f/\Delta$} & 
\colhead{$R_f/L_{\rm box}$}
}
\startdata
M01 & 7.61\ee7    &   50.4 &  1.53 & 0.00299 \cr
M02 & 2.53\ee8    &   75.2 &  2.28 & 0.00445 \cr
M03 & 8.42\ee8    &  112.3 &  3.40 & 0.00664 \cr
M04 & 2.80\ee9    &  167.6 &  5.08 & 0.00992 \cr
M05 & 9.32\ee9    &  250.2 &  7.58 & 0.0148  \cr
M06 & 3.10\ee{10} &  373.6 & 11.31 & 0.0221  \cr
M07 & 1.03\ee{11} &  557.8 & 16.90 & 0.0330  \cr
M08 & 3.43\ee{11} &  832.7 & 25.23 & 0.0493  \cr
M09 & 1.14\ee{12} & 1243.1 & 37.66 & 0.0736  \cr
M10 & 3.80\ee{12} & 1855.9 & 56.22 & 0.110   \cr
\enddata
\vskip-0.5in
\label{filtertab}
\end{deluxetable}

The method for generating these 11 grids is described in
great detail by \citet{martel05}.
Essentially, we work in $k$-space,
by generating, on a $512^3$ grid,
the density harmonics, $\hat\delta(k)$, 
corresponding to a $\Lambda$CDM power spectrum at $z=z_i=24$.
Once the density harmonics are generated we take the inverse Fourier
transform to obtain the initial density contrast, $\delta_i$. This gives us our
first grid in real space, with the unfiltered density contrast. 
To get the 10 filtered grids, we first
multiply the density harmonics by the Fourier transform of the filter,
and then take the inverse Fourier transform to get the filtered
density contrast. Notice that in our simulations
the value of the initial redshift, $z_i$, is actually 
irrelevant, as long as it is larger than the collapse redshifts of
all the density peaks that are resolved in our simulations. It
turns out that the first peak collapses at redshift, $z=18.55$.

\subsection{The Local Density Peaks}

For each of the 10 filtered grids, we identify the location of the
local density peaks. These peaks are defined as grid points where
the density contrast is positive, and its value exceeds the values at 
the 26 neighboring grid points (taking periodic boundary conditions into
account). For each density peak on each filtered grid, we compute a collapse
redshift, $z_{\rm coll}$, and a direction of least resistance, $\hat{\bf e}$.
The collapse redshift is obtained by solving numerically the following 
equation,
\begin{equation}
\label{zcoll}
\delta_c(z_{\rm coll})=\delta_i{\delta_+(z_{\rm coll})\over\delta_+(z_i)}\,,
\end{equation}

\noindent where $\delta_i$ is the initial density contrast at the peak,
and $\delta_+$ is the linear growing mode (for flat, $\Lambda\neq0$ models,
see, e.g., \citealt{martel91}, equation~[18]). The critical value,
$\delta_c$, is equal to 1.69 for an Einstein-de~Sitter model, and slowly varies
with redshift for other models. In our simulations, we simply assume
$\delta_c=1.69$. If equation~(\ref{zcoll}) has no solution, we are
simply dealing with a density peak that would collapse in the future,
and we just ignore it.

To check for consistency, we performed a N-body simulation of structure
formation in a box of the same size, for the same cosmological model
and density fluctuation power spectrum. We used a $\rm P^3M$ algorithm
\citep{he81} with $256^3$ equal-mass particles. The mass per particles
was $1.087\times10^7{\rm M}_\odot$, hence the lowest mass scale $M_1$
corresponds to 7 particles which, though insufficient for determining the halo
density profile, is fine for the simple locating of halos that follows.
 We used a standard friends-of-friends algorithm \citep{defw85}
to identify halos at various redshifts between $z=24$ and
$z=2$. For this, we used 2 different linking lengths. We first
use the ``standard'' value $l=0.25\Delta x$, where $\Delta x$ is the mean 
particle spacing. We also used a value 
$l=(18\pi^2)^{-1/3}\Delta x=0.1779\Delta x$, corresponding to a density
increase by a factor of $18\pi^2$. This is consistent with the assumption
that collapsed, virialized halos have a density equal to $18\pi^2$ times
the mean background density (see eq.~[\ref{tvir}] below). Only halos
containing 6 particles or more are included. This corresponds to a minimum
mass of $6.522\times10^7{\rm M}_\odot$.

\begin{figure}[t]
\hskip0.5in
\includegraphics[width=5.5in]{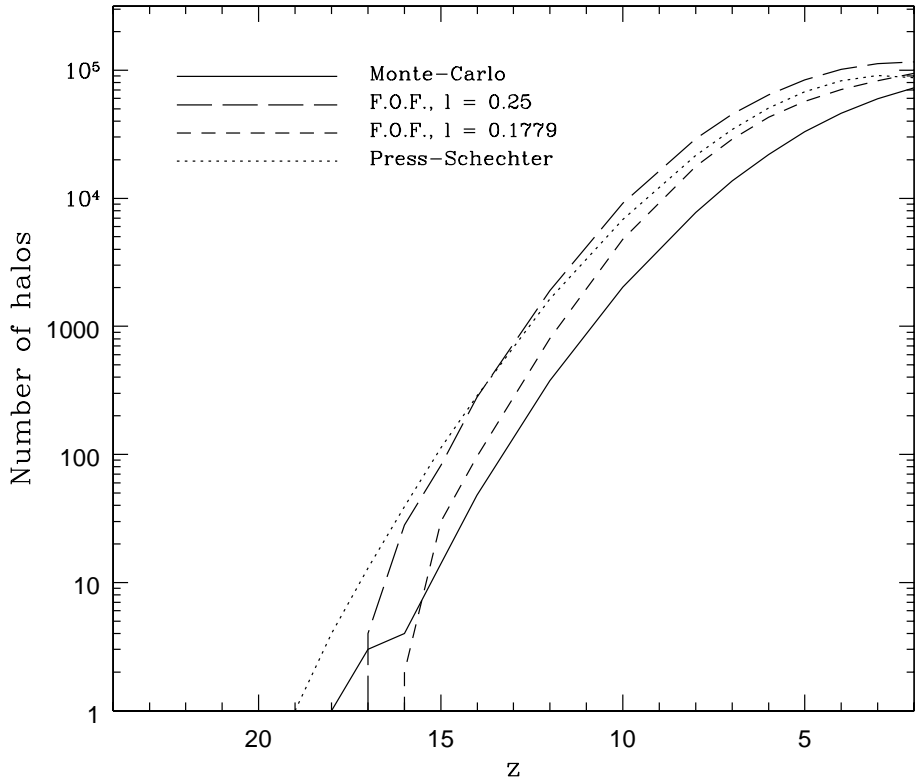}
\vskip-1.0in
\caption{Number of collapsed halos present in the computational volume,
versus redshift. {\it Long and short dashes}: N-body simulation with
friends-of-friends algorithm, with linking lengths, $l=0.25\Delta x$
and $l=0.1779\Delta x$, respectively; {\it dotted line}: Press-Schechter 
approximation;{\it  solid line}: Monte Carlo simulations.}
\label{fof}
\end{figure}

In Figure~\ref{fof}, we plot the number of collapsed halos
present in the N-body simulation
versus redshift (dashed lines). We find more halos with $l=0.25$ than
with $l=0.1779$. This might seem surprizing since halos ``break up'' as
$l$ is reduced. However, many halos fall below the threshold of 6 particles
when $l$ is reduced. The two methods differ by about 20\% at low redshift,
but the difference increases at high redshift. The dotted line shows a
calculation based on the Press-Schechter (PS) approximation \citep{ps74}. The
agreement with the N-body simulations is excellent at low redshift, while
at very high redshifts ($z>15$) the PS approximation overestimates
the number of halos. The solid line shows the number of collapsed
halos in our Monte Carlo simulations. This was obtained by counting, at
given redshifts $z$, the number of halos with $z_{\rm coll}>z$, and removing
the halos that have been destroyed by mergers (see \S3.4 below). The number
of halos is lower for the Monte Carlo simulations than for the N-body
simulation or the PS approximation, at almost all redshifts. 
We interpret this result as follows: in the Monte Carlo
simulations, only the matter located in overdense regions can eventually end
up inside collapsed halos, while in the N-body simulations, all the matter 
in the system can eventually end up in halos.
Notice also that the comparison between the PS and N-body results, and
the Monte Carlo results is complicated by the fact that the mass spectrum
of halos is discrete for the Monte Carlo simulations. For the N-body simulation
and the PS approximation, the choice of an appropriate minimum mass is not
obvious, and the results are quite sensitive to that choice, especially
at high redshift when most halos have a small mass.

These results also vindicate our choice of an initial redshift of, $z_i=24$.
Clearly the system is still in the linear regime at that redshift.
The very first halo collapses at redshift $z=18.55$ in the Monte Carlo
simulations, and at redshifts, $z=16-17$, in the N-body simulation,
while the PS approximation predicts that a computational volume
of that size will typically contain one halo at $z=19$. Indeed, the
probability of finding one halo of mass, $M=M_1=7.61\times10^7{\rm M}_\odot$,
at $z=24$, according to the PS approximation, is about 1/1000.

\subsection{The Direction of Least Resistance}

We compute, for each peak, the direction of least resistance,
using the method described in \S\ref{direction outflow}.
The radius, $R^*$, defines the region around the peak over which the
least-square fit to equation~(\ref{del1})
is performed. Its value will affect the determination of the coefficients,
$A$, $B$, $C$, $D$, $E$ and $F$, and ultimately the direction, $\hat{\bf e}$,
of the outflow. The largest possible value of $R^*$ 
should be the filtering scale, since this scale
corresponds to the physical extent of the peak (using a larger
value would amount to including matter that belongs to different peaks). 
As for the minimum value, $R^*$ must clearly be at least equal to the
grid spacing, $\Delta$. It turns out that this condition is insufficient.
If $R^*<2^{1/2}\Delta$, only the 6 nearest grid points to the peak are
included in the fit. In that case, the matrix, $M$, is already diagonal, and
the only directions allowed for the outflow are along the unrotated
coordinate axes $x$, $y$, $z$. We must at least include the 12 next nearest
grid points, located at a distance equal to $2^{1/2}\Delta$ from the peak.

\begin{figure}[t]
\hskip0.5in
\includegraphics[width=5.5in]{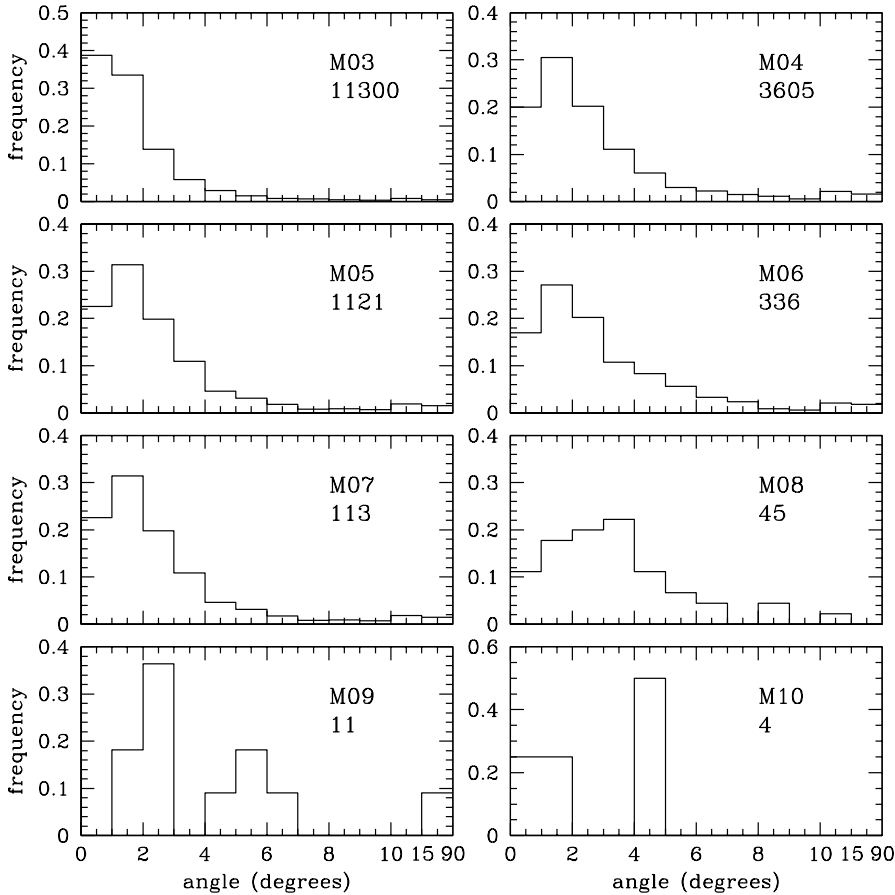}
\vskip-0.5in
\caption{Histograms of the fluctuations in the direction of the outflows
resulting from variations in $R^*$. Notice that the last two bins correspond
to the ranges [10,15] degrees and [15,90] degrees, respectively. The
labels in the panels indicate the mass filtering scale and the number of
peaks.}
\label{histo}
\end{figure}

To test how the direction of the outflow depends on
the particular choice of $R^*$,
we considered the filters, M03, M04, $\ldots$, M10, and
computed, for all peaks, two values of $\hat{\bf e}$: one using 
$R^*=2\Delta$, and one using $R^*$ equal to the filtering scale, $R_f$.
Figure~\ref{histo} shows histograms of the angle in degrees between these two
values of $\hat{\bf e}$. 
We excluded from this test the mass filters
M01 and M02, because the range of allowed values for $R^*$ 
becomes very narrow, and there is essentially no uncertainty in these cases. 

For the vast majority of peaks, the angle is
below 4 degrees and it exceeds 10 degrees only for very few peaks.
The interpretation of these results is that our method for determining
the direction of the outflow has an uncertainty of a few degrees, resulting
from the ambiguity in the choice of $R^*$. 
In the limit when the opening angle of the outflow is much larger
than that uncertainty, the region of space containing the outflow
is essentially unchanged. We will consider opening angles of several 
tens of degrees, while the uncertainty
in the direction of least resistance is of order a few degrees. The final results might vary by
 a few percent as a result, but this is not  sufficiently large to be the dominant 
 source of error.

From this test, we conclude that our technique for determining the
direction of least resistance is robust, and consequently the
particular value chosen for $R^*$ is not important. Hence, we
decided to set $R^*$ equal to the filtering scale, $R_f$.

Once this is all done, we have, for each of the 10 filtering scales,
a list of local density peaks, and for each of these peaks, a position, 
${\bf r}$, a mass, $M$ (the mass of the corresponding filtering scale), 
a collapse redshift, $z_{\rm coll}$, and a direction of least
resistance, $\hat{\bf e}$. In the model, each peak corresponds to a density
fluctuation of mass, $M$ (the filtering mass), and radius, $R_f$ (the
filtering radius). If left alone, this peak will collapse
at redshift, $z_{\rm coll}$, to form a halo of mass, $M$, 
located at position, ${\bf r}$. This protogalaxy will eventually turn into
a galaxy (by forming stars), and generate an outflow propagating along the
direction of least resistance, $\hat{\bf e}$.

\subsection{Formation of First-Generation Galaxies}

The 10 filtered grids represent the same region of the universe. Hence,
we cannot simply assume that every peak collapses to form a protogalaxy,
since this would violate conservation of mass. We must deal with the issue
of halos inside halos.\footnote{Throughout this paper,
we shall use the term ``halo'' to designate a collapsed object,
whether it is a galaxy or a protogalaxy. } 
Consider two density peaks $a$ and $b$ of
masses, $M_a<M_b$ (these peaks are therefore on different grids). If the
separation, $|{\bf r}_a-{\bf r}_b|$, between the peaks is smaller than the
filtering scale, $R_{f,b}$, of the larger mass, the two halos that these peaks
will form cannot coexist, since the smaller halo is inside the larger one.
There are then two possibilities, depending
on the collapse redshifts. If $(z_{\rm coll})_b>(z_{\rm coll})_a$,
then halo $a$ will never form, since all the matter it contains
will be incorporated into halo $b$, that forms earlier. While this does occur
it is not common, since in a CDM universe smaller
peaks tend to collapse earlier than bigger ones. The second possibility
is $(z_{\rm coll})_a>(z_{\rm coll})_b$. In this case, halo $a$ will form
first, at redshift $(z_{\rm coll})_a$, but will be destroyed later
by a {\it merger event} at redshift, $(z_{\rm coll})_b$, that results in
the formation of halo $b$. So, even though halos remain at
fixed locations in this simplified Monte Carlo model, non-dynamic merger events
are included.

The collapse of a peak leads to the formation of a halo,
which we identify as a protogalaxy. The gas inside that halo is at the
virial temperature, given by
\begin{equation}
\label{tvir}
T_{\rm vir}={0.009\,{\rm K}\over\beta}\left({6.8\over5X+3}\right)
\left({Mh\over1\,{\rm M}_\odot}\right)^{2/3}(1+z_{\rm coll})
\left[{\Omega_0\over\Omega(z_{\rm coll})}\right]^{1/3}
\left[{\Delta_c(z_{\rm coll})\over18\pi^2}\right]^{1/3}\,,
\end{equation}

\noindent \citep{ecf96}, where $X=0.76$ is the hydrogen mass fraction,
and $\Delta_c$ is the ratio of the halo's mean density to the
critical density. Since this temperature is usually much too high
to allow the formation of stars, the gas must cool down to a temperature,
$T\ll T_{\rm vir}$, before stars form and the protogalaxy becomes a galaxy.
As in SB01, we use the cooling model of \citet{wf91}. This model assumes that
the cooling proceeds inside-out. The gas inside a ``cooling radius,'' 
$r_{\rm cool}$, has cooled, while the gas outside $r_{\rm cool}$ remains hot.
As $r_{\rm cool}$ increases with time, the mass, $M_{\rm cool}$, of cool
gas increases, until all the gas has cooled. The mass, $M_{\rm cool}$,
evolves according to
\begin{equation}
\label{dmcooldt}
{dM_{\rm cool}\over dt}=12\left({\Omega_{b,0}\over\Omega_0}\right)^{3/2}f_0
\left({T_{\rm vir}\over1\,{\rm K}}\right)
\left[\Lambda_{23}(T_{\rm vir},Z)\right]^{1/2}
\left({t\over 1\,{\rm yr}}\right)^{-1/2}M_\odot\,{\rm yr}^{-1}\,,
\end{equation}

\noindent where $\Lambda_{23}$ is the cooling rate in units of
${\rm 10^{-23} ergs\,s^{-1}cm^{-3}}$, which is a function of temperature and
metallicity, $Z$, and $f_0\approx0.8$ [a correction factor 
suggested by \citet{s97}
to bring this result into line with more complex analyses]. 
We determine the required cooling rate 
using MAPPINGS III\footnote{http://www.mso.anu.edu.au/$\sim$ralph/map.html}, a
successor to MAPPINGS II outlined in \citet{sd93}.
We integrate both sides between $t=0$ and
$t=t_{\rm cool}$, the cooling time. For a halo of mass, $M$, and gas mass,
$M_{\rm gas}=\Omega_{b,0}M/\Omega_0$, we get
\begin{equation}
t_{\rm cool}={1\over576f_0^2}
\left({M\over1{\rm M}_\odot\,}\right)^2
\left({\Omega_{b,0}\over\Omega_0}\right)^{-1}
\left({T_{\rm vir}\over1\,{\rm K}}\right)^{-2}
\left[\Lambda_{23}(T_{\rm vir},Z)\right]^{-1}
{\rm yr}\,.
\end{equation}

Using the solution of the Friedman equation for our $\Lambda$CDM
model, we compute, for each halo, the epoch, $t_{\rm coll}$, 
at which collapse occurs from the
collapse redshift, $z_{\rm coll}$. We then
add the cooling time to get the epoch at which galaxy
formation occurs, $t_{\rm gf}=t_{\rm coll}+t_{\rm cool}$,
and compute
the corresponding galaxy formation redshift, $z_{\rm gf}$. The galaxy formation
epoch may be later than the current epoch, in which case
stars will not form and the halo will remain a protogalaxy, unless metal
deposition from outflows modifies the cooling rate (see below).

We refer to these galaxies as ``first-generation'' 
galaxies since at the time of
formation they are untouched by impinging outflows from other galaxies. For 
these galaxies we assume a cooling rate based on zero metallicity. In 
the following section we will discuss galaxies for which this is not the case.

We are now ready to proceed with the simulation. Starting at initial redshift
$z_i=20$ (since there are no collapsed halos at higher redshift), 
we evolve the system forward by redshift steps of $\Delta z=-0.005$. 
As a result we take 3600 steps to reach the end point for our simulations, 
which is
$z=2$. This end point is chosen for comparison with Quasar absorption 
spectra data from $z=2-6$. Also, our box would no longer represent a
cosmological volume at lower redshifts.

At the location of peak, $a$, a halo forms at redshift, 
$(z_{\rm coll})_a$, unless that peak was located inside a larger peak that
collapsed first, as described above. During the redshift interval, 
$[(z_{\rm coll})_a,(z_{\rm gf})_a]$, while the gas is cooling, the halo might
be destroyed by a merger. This will happen if there is a peak $b$ for which
\begin{eqnarray}
&&M_b>M_a\,,\\
&&(z_{\rm coll})_a>(z_{\rm coll})_b>(z_{\rm gf})_a\,,\\
&&|{\bf r}_a-{\bf r}_b|<R_{f,b}\,.
\end{eqnarray}

\noindent In this case, the halo was able to form, but did not manage
to cool before being destroyed, hence the actual galaxy never
formed. 

If the galaxy does manage to form, we neglect the formation time and life span
of massive stars and so the newly formed galaxy immediately produces
an outflow centered at
the halo position, ${\bf r}_a$, and propagating along the direction, 
$\hat{\bf e}_a$. We evolve this outflow with time, using the solutions given
in Appendix A (with timesteps corresponding to our redshift 
steps, $\Delta z=-0.005$).

Left alone, the outflow will go through successive phases of
driven expansion by supernovae and near-adiabatic expansion driven by pressure alone 
before joining the Hubble flow. However,
during either of these stages, the galaxy producing the outflow might be
destroyed by a merger. In all such cases the outflow reverts to a Hubble 
expansion phase.

\subsection{Subsequent Galaxy Formation}

After the first outflows are formed, some will begin to strike other 
density peaks\footnote{
In the following text we will use the term ``peaks'' to refer to 
density peaks that will go on 
to form halos by $z=0$. } 
 and modify the way these systems evolve. They may enrich these 
systems with new metals, possibly modifying their cooling time.
They may also expel the gas content of the object by ram pressure stripping
or shock-heating of the gas. 
Of these two mechanisms, the former (stripping) is the 
dominant one \citep{sfb00} and so we will neglect the latter. 
 
\subsubsection{Stripping}

Density peaks may be stripped of their baryons when the swept-up shell 
of intergalactic and interstellar gas incident upon them 
imparts sufficient momentum on the baryons such that they escape the 
potential well of the peak, i.e.
\begin{equation}
\left({l^2 \over 4R^2}\right) M_ov_o \ge M_b v_{\rm esc},
\end{equation}

\noindent
where the mass of the swept up shell is $M_o$ at radius, $R$, (as in TSE), 
$v_o$ is the outflow velocity, $M_b$ is the baryonic mass of the
density peak
struck, and $v_{\rm esc}$ is the escape velocity for successfully 
stripped gas.
$l$ is the comoving radius of the collapsing density peak. Using the spherical 
approximation for gravitational collapse, this scale is determined from the 
non-linear density contrast, $\delta_{\rm NL}$ 
using $l=R_f(1+\delta_{\rm NL})^{-1/3}$ where
$1+\delta_{\rm NL}= 9(\theta-\sin \theta)^2/[2(1-\cos\theta)^3]$. 
The parameter, $\theta$, is in
turn given by 
\begin{equation}
\left(\theta-\sin\theta\over2\pi\right)^{2/3}= 
{\delta_+(z)\over\delta_+(z_{\rm coll})}\,,
\end{equation}

\noindent
where $\delta_+(z_{\rm coll})$ and 
$\delta_+(z)$ are the linear growing modes at collapse and 
when the outflow strikes the peak respectively.

We assume that stripping may only occur for systems that have not collapsed 
since the cross section to impact is small and would result in 
negligible momentum 
transfer. This approach is supported by \citet{sfs05} who 
investigate this issue 
with a numerical analysis of an individual object struck by a shock.
It is also 
worth noting that we deal with baryonic stripping in the same way whether 
or not the source galaxy is 
within the filtering radius of the density peak struck (although it is 
less likely to occur due to the 
larger escape velocity and mass of the halo struck).

When the criterion for stripping is met, the density peak is rendered 
free of baryons and, while it may collapse, it will not form a galaxy. 
If the shell of the outflow does not strip the peak then the peak is
enriched by the metal-rich gas, which fills the outflow.

\subsubsection{Metal Enrichment}

Metals are propagated throughout our simulations within the hot bubbles of 
outflows.
Peaks that have not been struck by outflows are assumed to have metallicity  
of $\rm[Fe/H]=-3$, which is negligible for the purposes of 
calculating the halo 
cooling time. Once galaxies are formed metals are produced at rate of 
$2{\rm M_\odot}$ per SN \citep{ns98}. Hence the mass of metals in the 
outflow is
\begin{equation}
M_Z=f_{\rm esc}{2{\rm M_\odot}\over M_{\rm req}}f_*{\Omega_{b,0}\over\Omega_0}
M\,,
\end{equation}

\noindent
where $f_{\rm esc}$ is the fraction of ISM gas 
blown out with the outflow. We use the 
value $f_{\rm esc}=0.5$ taken from the numerical simulations of \cite{mfm02}. 
This mass of metals is distributed evenly throughout the volume of the outflow.

It is notable that, in this model, the ratio of the mass lost rate
to star formation rate is approximately $f_{\rm esc}/f_*=5$.
This is consistent with observations that
indicate this ratio is of order 1 for local starburst galaxies 
(e.g. \citealt{martin99} and \citealt{heckmanetal00}), infrared-luminous 
galaxies 
(e.g. \citealt{rupkeetal05}) and high-$z$ galaxies 
(e.g. \citealt{pettinietal00}).

When an outflow strikes an uncollapsed density peak and
does not strip it, it modifies its metal 
content. It deposits a fraction of its metals, 
$f_{\rm dep}V_{\rm overlap}/V_{\rm outflow}$, where 
$f_{\rm dep}$ is a mass deposition 
efficiency, which we set at $f_{\rm dep}=0.9$, 
$V_{\rm overlap}$ is the volume of overlap of the uncollapsed 
density peak of radius $l$, and  $V_{\rm outflow}$ 
is the volume of the outflow. 
This volume of overlap is calculated based on various geometric approximations 
to the overlap of two spherical cones and a sphere, depending on 
their relative 
size and orientation. Once the halo has collapsed, the addition of metals
results in an increase in cooling rate and so a reduction in the cooling 
time, expediting galaxy formation.

We assume that the SNe in our simulations are only of the Type~II variety
since these SNe explode together shortly after the starburst and lead to
a coherent extragalactic outflow that will reach large distances.
As a result we use an alpha-element-rich yield of metals [quoted in
\citet{sd93} as ``primordial'' abundance ratios] and distribute our 
mass of metals accordingly. This provides a value for ${\rm [Fe/H]}$ used to
calculate the new cooling rate.

There is a special case to consider where a source galaxy is embedded in 
the density peak that its outflow strikes. In this case we assume the volume
overlap of the outflow is 100\% and so it will deposit a fraction, $f_{\rm dep}$, 
of its metals onto this structure. The main motivating 
factor behind the inclusion of this mass deposition efficiency is to avoid the 
implausible scenario that an outflow may continue to grow having left all 
its metals behind. The value it should take is unknown; however,  the 
simulations are not sensitive to this factor, as we point out below.

\section{RESULTS}

\subsection{The Impact of Varying Opening Angle}

We have constructed a model for the evolution of anisotropic outflows and a 
test bed with which to investigate its significance and we are now ready to 
discuss what this approach tells us.

\begin{figure}[ht]
\vskip-0.1in
\hskip-0.5in
\includegraphics[width=8in]{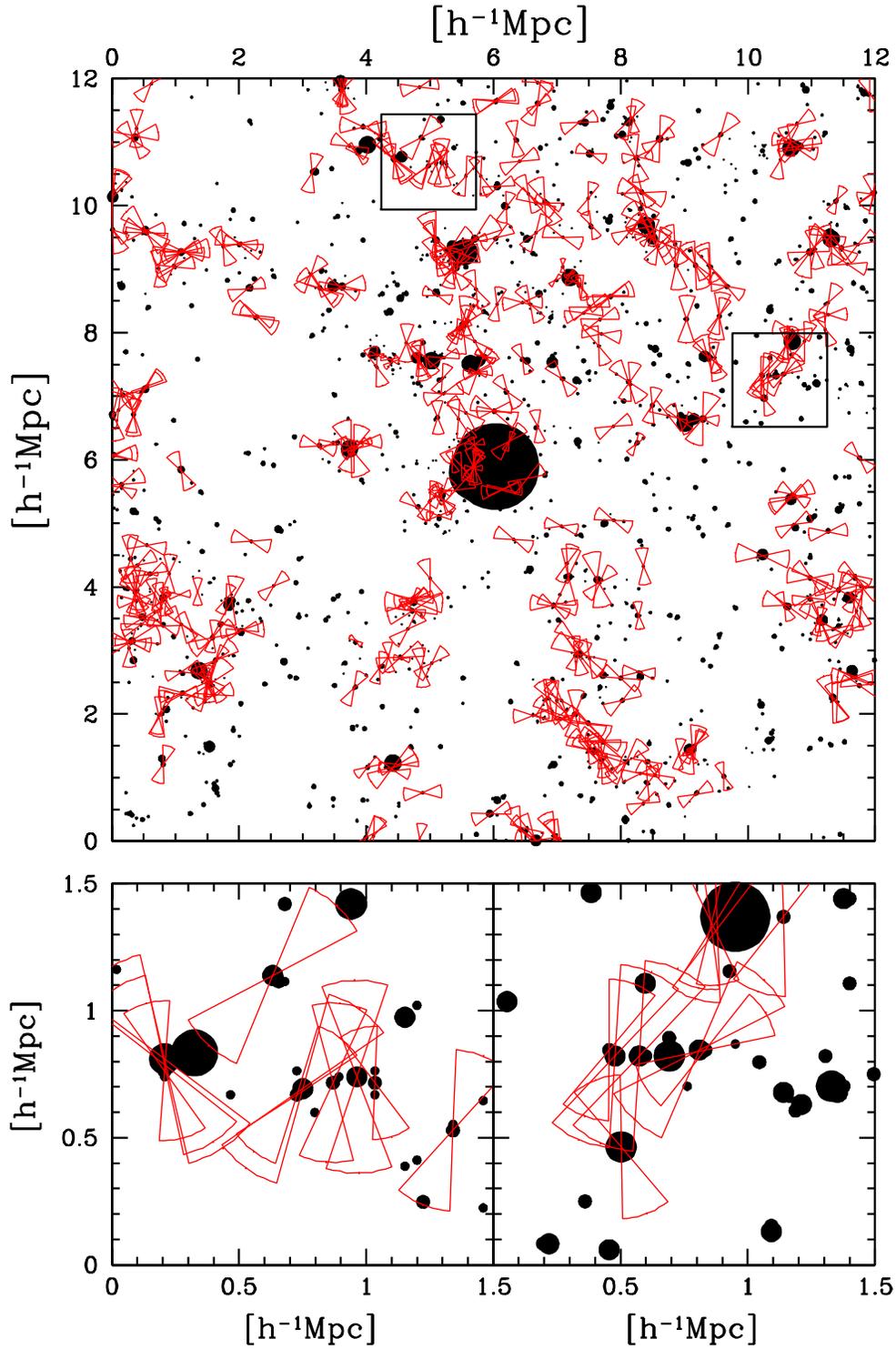}
\vskip-0.2in
\caption{A slice through the simulation box of thickness  
$0.4h^{-1}{\rm Mpc}$ at $z=2$.
Density peaks are shown as filled circles of diameter corresponding to their 
extent prior to 
collapse. The outflows have an opening angle of $40^\circ$, and their 
location and physical coverage are indicated as red 
wedges. Two zoom-in regions of size 
$1.5h^{-1}{\rm Mpc}$ show regions of interest,
where galaxies that formed out of a common filament produce
outflows that are aligned.}
\label{slice}
\end{figure}

Taking an extreme case of anisotropy as an illustration, consider an opening 
angle of only $40^\circ$. At the natural end point of our simulations, $z=2$, 
 Figure~\ref{slice} shows a slice of thickness $0.4h^{-1}{\rm Mpc}$ 
through our simulation box. The position and extent of our galactic outflows
are indicated along with the 
position and pre-collapse radius of peaks (i.e. the filtering scale, $R_f$, of
the peak) that will collapse by $z=0$. These outflows
extend a significant distance from their source galaxies and often strike 
other peaks. Where peaks are arranged in a row in the plane of the slice, 
the outflows appear to favor a direction perpendicular to this structure 
(see zoom in). This is to be expected: the locations of these galaxies trace
the underlying structure of the dense filament out of which they form by 
fragmentation, and the outflows follow the path of least resistance away from 
that filament.

\begin{figure}[t]
\hskip0.5in
\includegraphics[width=5.5in]{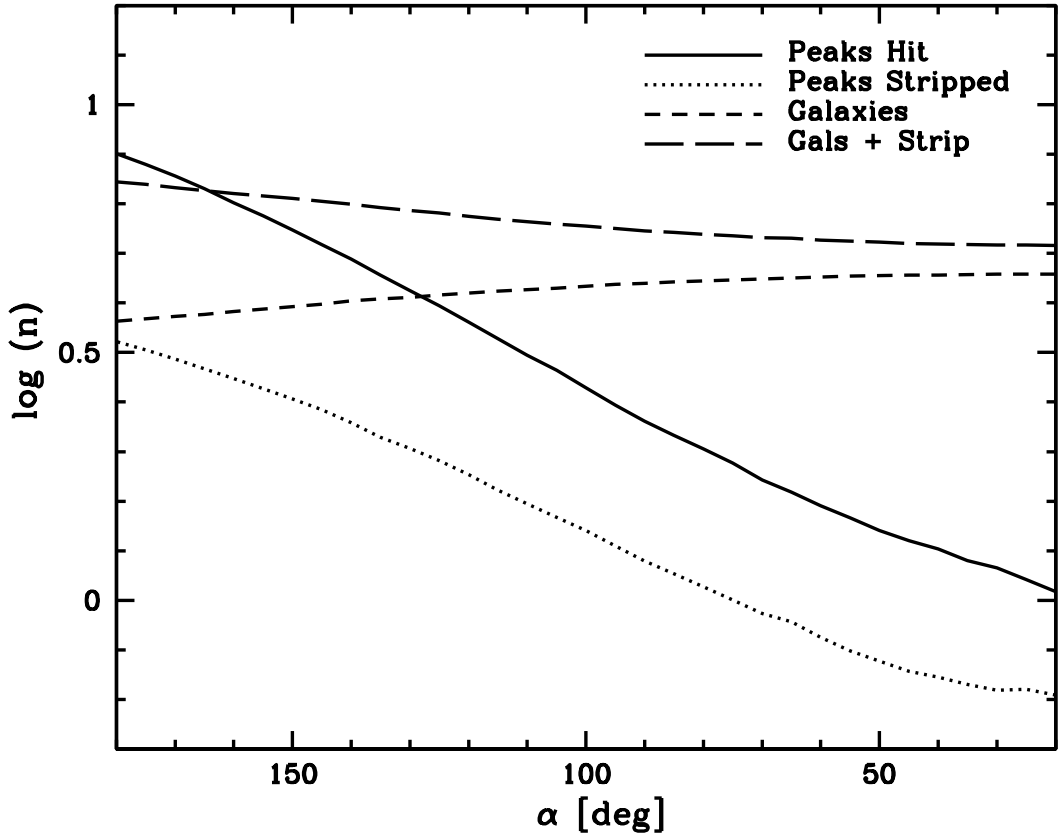}
\caption{ The number density of various objects in units 
$(h^{-1}{\rm Mpc})^{-3}$ at $z=2$, versus opening angle.
{\it Solid line}: number of peaks hit that will go on to 
collapse before $z=0$; {\it dotted line}:
number of those peaks that are then stripped of their baryons; 
{\it short dashed line}: number of galaxies;
{\it long dashed line}:
sum of the number of galaxies and stripped peaks.
} \label{counts}
\end{figure}

\begin{figure}[t]
\hskip0.2in
\includegraphics[width=6.0in]{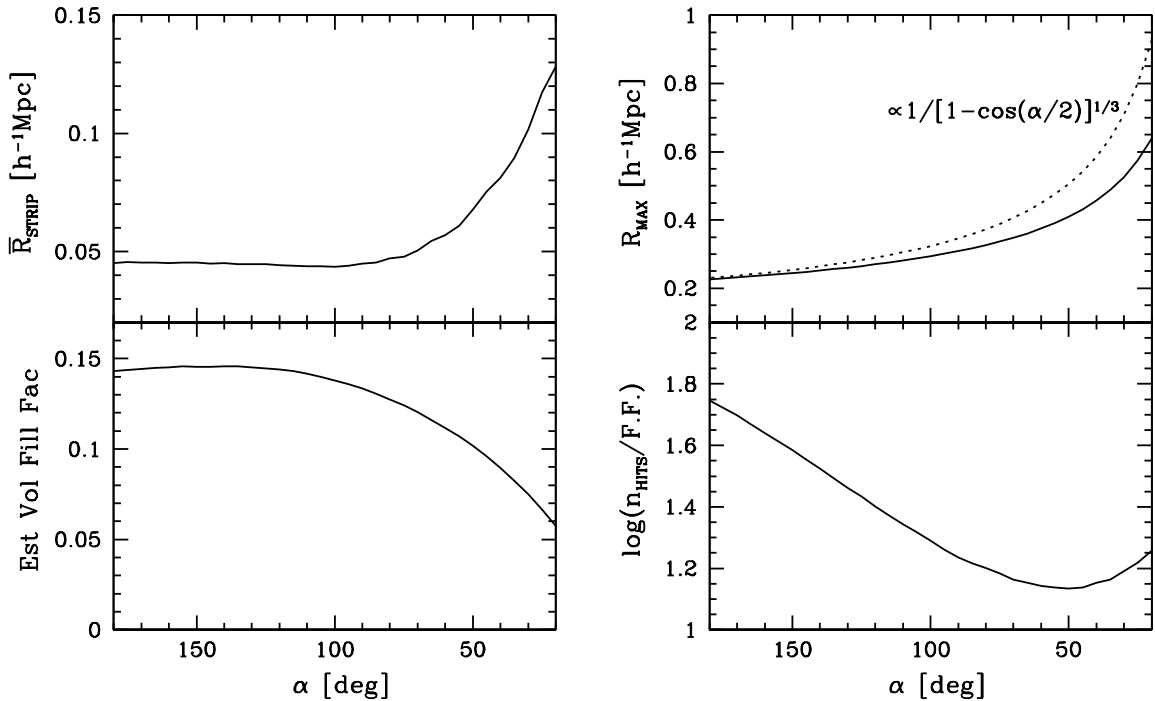}
\caption{{\it Top left}: mean radius of outflows when they 
strip peaks of their baryons; {\it top right}: the {\it solid line} 
shows the radius of the largest outflow and, for comparison, the 
{\it dotted line} 
shows $R\propto1/[1-\cos{\alpha/2}]^{1/3}$, which is the relation 
expected if the volume enriched were conserved for varying opening angle. The 
estimated volume filling factor is shown in the {\it bottom left} panel;
{\it bottom right}: the ratio of the number density of peaks 
hit to the volume filling factor in units of $(h^{-1}{\rm Mpc})^{-3}$.
} \label{endstats}
\end{figure}

Figure~\ref{counts} shows the counts of various quantities in units 
of $(h^{-1}{\rm Mpc})^{-3}$ by $z=2$, for varying opening angle. The number 
density of peaks that are hit by outflows that will go on to form halos 
by $z=0$ are shown. This decreases as approximately a power law 
for increasingly anisotropic outflows, which can be explained by two factors. 
Anisotropic outflows take a path of least resistance out of highly overdense 
regions and so will encounter fewer regions that go on to form halos. Also the 
total volume occupied by outflows is not conserved for varying opening angle 
since anisotropic outflows tend to overlap less (see above) and the volume 
per outflow is not constant. The degeneracy between these two effects is 
resolved in Figure~\ref{endstats} and the accompanying text. The number 
density of peaks stripped of their baryons is also shown 
and indicates that the fraction of peaks hit that are then
stripped is roughly constant ($\sim 50\%$).

The number of galaxies formed by $z=2$ rises smoothly for decreasing opening 
angle in Figure~\ref{counts} as a result of the fall in stripping. We find 
that metal enrichment of peaks has a negligible impact on the number of 
galaxies; however, those galaxies that form are born slightly earlier. Not 
all peaks that are hit by outflows but not stripped will go on to produce 
galaxies by $z=2$. This is because, despite being enriched by 
metals and collapsing,
their baryons do not always cool to form stars by this redshift. As a 
result, the incidence 
of stripping decreases faster than the number of galaxies 
increases. This is demonstrated with the sum of these two number densities 
in Figure~\ref{counts}, which decreases with increasingly anisotropic outflows.

\clearpage

In Figure~\ref{endstats} we show various statistics at $z=2$ in the 
simulations: the average
distance, $\bar R_{\rm strip}$,
traveled by outflows before stripping occurs, 
the maximum distance, $R_{\max}$, 
traveled by an outflow, the estimated volume filling factor of outflows, and 
the ratio of the number density of hits to the volume filling factor in units 
$(h^{-1}{\rm Mpc})^{-3}$. The mean distance traveled by outflows before 
stripping occurs is flat until the opening angle as low as $100^\circ$ 
where it begins to increase. 
As the anisotropic outflows travel farther with decreasing opening 
angle, they fail to hit new peaks as they have already 
escaped their highly overdense environments. It is only when 
$\alpha=100^\circ$ that they begin to reach the next dense structure.

In the top right panel we show the radius of the largest outflow for each 
opening angle. We also show a curve for constant volume 
using equation~(\ref{volume}),
which gives us $R\propto1/[1-\cos{(\alpha/2)}]^{1/3}$. 
This shows that while the radii of 
outflows grow with decreasing opening angle they do not grow fast enough 
to conserve the volume per outflow. 

The estimate of the volume filling factor is calculated based on the sum of 
the volume of all outflows, with a correction factor for the increasing 
probability of overlap for large values,
\begin{equation}
\label{volestimate}
F=1-\exp\left[- {\sum^{N_o} _{i=1} V_i \over (12h^{-1}{\rm Mpc})^3}\right],
\end{equation}
where $V_i$ is the volume per outflow given by equation~(\ref{volume}) and 
$N_o$ is the number of outflows.
The volume filling factor rises to a peak at $\alpha\sim140^\circ$ but is essentially 
constant from $\alpha=110^\circ-180^\circ$. As 
indicated above, the volume per outflow is not conserved with varying opening 
angle and the volume of the enriched region actually decreases for 
increasingly 
anisotropic outflows. As a result one would naively expect the volume filling 
factor to fall for decreasing $\alpha$; however, since the number of galaxies 
increases the volume filling factor holds constant until $\alpha\sim110^\circ$ 

Figure~\ref{counts} indicates that the number density of peaks hit decreases 
from $180^\circ-20^\circ$, but so does the volume filling factor 
(Fig.~\ref{endstats}). We can use 
both these statistics to determine the impact of varying $alpha$ on the change 
in the number of peaks hit resulting from the path of least resistance of 
the outflows. We do this by taking the ratio of the number density of hits 
to the volume filling factor, shown in Figure~\ref{endstats}; this is 
the number of hits per volume covered by outflows. This indicates that as 
outflows become more anisotropic they tend to avoid high-density structures 
and favor voids until $\alpha \approx50^\circ$ when this statistic 
turns around,
indicating that most outflows have crossed voids and have reached the 
next overdense structure. 

\begin{figure}[ht]
\hskip0.9in
\includegraphics[width=4.8in,angle=90]{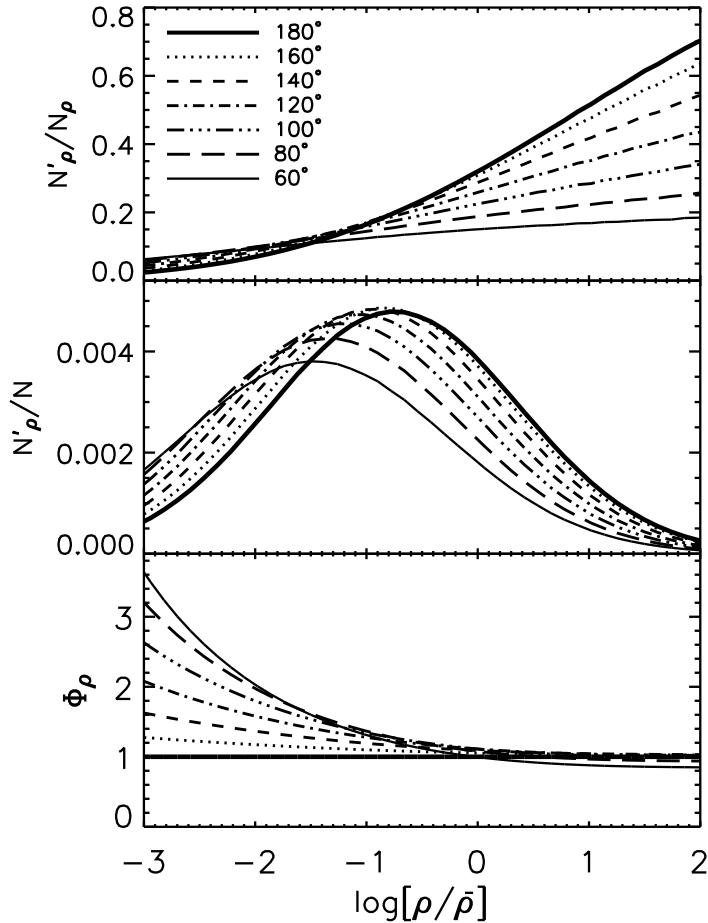}
\caption{The number of enriched grid points $N^\prime_\rho$ in the 
simulation volume at $z=2$ 
as a function of IGM overdensity for varying opening angle. In the {\it top} 
panel this 
is shown as a fraction of the number of systems at this density, $N_\rho$. 
The statistic as a function of the total number of points, $N$, is shown in 
the {\it middle} panel. The {\it bottom} panel shows the number of enriched 
points below an overdensity threshold, $N^\prime_{\rho^\prime<\rho}$ by
comparison with the isotropic case using $\Phi_\rho$ given in equation~(29).
} \label{enrich}
\end{figure}

These statistics indicate that galactic outflows 
undergo a transition over the range $\alpha=100^\circ-50^\circ$, 
from enriching their 
own high-density sources and the surrounding voids, to enriching neighboring 
high-density regions after crossing the voids. The top right panel indicates 
that in this range the radius of the largest outflow increases from around 
$0.3-0.4h^{-1}{\rm Mpc}$ compared to the isotropic case of 
$0.23h^{-1}{\rm Mpc}$, confirming the plausibility of this explanation.

\subsection{Enrichment of Overdense and Underdense Systems}

We have established that the volume filling factor of enriched regions is
dependent on the anisotropy of galactic outflows, but what is the nature 
of regions enriched? We investigate this by returning to the original power spectrum,
and smoothing on a Jeans length scale to determine the baryonic density field 
produced with the same Gaussian random realization 
of structure. This density field is a result of purely linear evolution from 
initial conditions with a Gaussian probability distribution function (PDF). We
map to a lognormal PDF in order to mimic a degree of non-linear 
behavior. This has been found to reproduce the PDF of the density 
in SPH simulations \citep {bd97} but is limited by an incomplete description
 of clustering. 

Having scanned  through all $512^3$ grid points in the mapped density field,
we determined whether any of the galactic outflows reach them. Where this is 
the case we have flagged them as enriched and noted their overdensity. In 
Figure~\ref{enrich} we show three statistics
derived from this approach for a range of different opening angles from 
$60^\circ-180^\circ$. We show the number of grid points 
enriched at a given overdensity, $N^\prime_\rho$, as a fraction of the number 
of grid points at that density, $N_\rho$  (top panel); this is effectively the probability 
of enriching a systems of a given density. This panel indicates that the 
effect of galactic outflows on overdense systems is dramatically reduced for 
increasingly anisotropic outflows. This is quite plausible in the context of 
the low number density of overdense systems. It is also clear that for 
decreasing opening angle the probability of enriching low-density systems 
increases.

The middle panel shows the number of grid points enriched at a given 
overdensity as a fraction of the total number of grid points, $N$. In a plot 
of $\log(\rho/{\bar \rho})$ these curves are
Gaussian reflecting the underlying density PDF. The mean of the Gaussian, 
drifts to lower densities for lower opening angles. Even in the case of 
isotropic outflows the majority of metals
are in underdense regions when considered by volume and as the outflows
become anisotropic this effect becomes stronger as winds expand 
preferentially into low-density regions.

The bottom panel is a cumulative version of the middle 
panel and so indicates the number of enriched grid points with an overdensity 
below a given threshold, $N^\prime_{\rho^\prime<\rho}$.
This statistic is shown by comparing to the isotropic case using
\begin{equation}
\Phi_\rho = N^\prime_{\rho^\prime<\rho}/N^\prime_{\rho^\prime<\rho, 180}.
\end{equation}
 This panel highlights the significant impact on 
the enrichment of underdense systems of anisotropic outflows. Anisotropic 
outflows can lead to an increase in the enriched volume of underdense systems 
of $10\%$ (where $\alpha=100^\circ- 120^\circ$) and an increase of $40\%$
in systems below $\rho/{\bar \rho}=0.1$ (where $\alpha=80^\circ- 100^\circ$) 
 compared to isotropic outflows. The real volume filling factor can be determined 
 and reaches a maximum of $11\%-12\%$ for $\alpha\approx140^\circ$.

\section{DISCUSSION}
 
There is a degree of uncertainty for the parameters $f_*$, $f_{\rm w}$, 
$f_{\rm dep}$,  $f_{\rm esc}$ and they are of varying importance. The star
formation efficiency,  $f_*$, has a well-documented and 
large uncertainty. It has
a significant affect on the predicted volume filling factor as 
outlined in SFM.
Our quoted values for the volume filling factor should be considered with this
in mind. However, the dependence on opening angle will remain unchanged.
The fraction of total supernova energy that contributes to
 the outflow, $f_{\rm w}$, and the mass escape fraction, $f_{\rm esc}$, both have
  associated uncertainties. These uncertainties have been theoretically 
  constrained by SFM (and references therein) and \citet{mfm02}. We are
   aware of no direct observational constraints; although our ratio of $f_{\rm esc}/f_*$
  is consistent with observation by \cite{martin99}, \cite{heckmanetal00}, 
\cite{pettinietal00}, \cite{rupkeetal05} and others.

 The value for the efficiency with which
outflow metals are deposited upon collapsing density peaks, $f_{\rm dep}$,
is poorly constrained and requires further analysis. This is not a concern here 
as the metal enrichment of halos does not have a significant impact on the subsequent
numbers of galaxies and volume of the IGM enriched. Only a handful of 
additional galaxies form before $z=2$, with the inclusion of increased
 cooling by metal enrichment, that would otherwise not have formed (i.e. if
 $f_{\rm dep}=0$). All density peaks that are enriched, and that will collapse 
 to form galaxies in our simulations, will form earlier. This raises the distance 
 traveled by outflows whose expansion is halted either by merger of the source
  galaxy or the end of the simulation run. The volume enriched falls by less than 
  $0.1\%$, when reducing the deposition efficiency from $f_{\rm dep}=0.9$ to
  $f_{\rm dep}=0$.

Where the outflows stall before their radii are larger than the pre-collapse
scale of their density peaks, we can reasonably assume that the outflows 
will fall back and galactic fountains will be formed. Since we allow all our 
outflows to join the Hubble flow when the internal pressure equals the 
external pressure, we have not attempted to simulate this effect. This will
not change our results to a significant degree as this is the case in a small
minority of our outflows. Even in those few cases, no stripping or metal 
enrichment of other peaks occur and they fill only a small fraction of the 
universe so making little impact on the volume filling factor statistics shown.

In the calculation of external pressure we assume helium is singly-ionized, 
which is true for most of the redshift range where the outflows affect the IGM 
and will have little effect on the pressure.
We also assume that the IGM is at the mean density and temperature
and that the mean temperature is not redshift dependent. The density of the 
IGM is also implicitly assumed to be constant in the second term of 
equation (4) corresponding to the drag due to the sweeping up of
the IGM (see TSE for more details).
Most of the enriched volume is in underdense regions for all opening angles 
considered as 
shown in Figure 9 (discussed below) and the temperature of the IGM will be lower than 
that at mean density. Both these factors lower the external pressure, and
 so we can expect outflows to travel further after this correction. 
 Outflows that only travel a short distance, such that they do not escape overdense
 regions, will travel even shorter distances as a result. A further 
 complication arises since the outflows themselves will raise the temperature
 of the IGM. We will seek to investigate these issues in future work.

We assume that the IGM is ionized in our calculation of the Compton drag on
expanding outflows. This appears to be reasonable since recent {\sl WMAP} results
indicate that the epoch of reionization is $z\sim11$ \citep{p06} while outflows before 
$z\sim9$ have a negligible impact on our results.
 
The results from the previous section consistently indicate that we 
successfully describe anisotropic outflows that take a path of least 
resistance out of overdense regions, such as pancakes and filaments, 
and into voids. If sufficiently anisotropic 
($\alpha \lesssim 100^\circ$) these outflows will also begin to strike 
neighboring overdense structures and further peaks. 
The number of peaks hit drops for anisotropic outflows and diminishes the 
ability of these outflows to strip peaks of their baryons. This reduces the 
capacity of ram pressure stripping at high-$z$ to explain the abundance of dwarf 
galaxies in the Local Group.

We find a value for the volume filling factor of $11\%-12\%$ 
 (for $\alpha\approx140^\circ$), but this 
is sensitive to a number of factors such as the star formation efficiency 
and the degree of clustering of source galaxies.
We will present a more thorough analysis of the value 
of this quantity in future papers in this series. We do, however, expect its 
dependence on opening angle of outflows to be rigorous. 

Despite this small volume filling factor, we find  
significant enrichment of underdense systems
(below the mean density of the Universe),
particularly when these 
outflows are anisotropic. The volume filling factor of enriched underdense 
systems is maximized by an opening angle of $100^\circ-120^\circ$, this is 
$\sim 10\%$ higher than that for isotropic outflows, while the volume of enriched 
space below $\rho/{\bar \rho}=0.1$ is $40\%$ higher. 
This may provide an explanation for 
observations (at high-$z$) of metal enrichment in systems at, or around, the mean density as 
seen in \citet{2003ApJ...596..768S} and \citet{2004MNRAS.347..985P} without 
the need to appeal to large volume filling factors. These outflows also 
reach larger distances from their source galaxies and help explain
why metal enrichment is seen far from observed galaxies at high redshift
\citep{psa06, s06}, while still showing elevated metal enrichment 
close to those galaxies
\citep{a03,a05,psa06, s06}. Since parts of the IGM are enriched solely 
based on whether 
they are located in the path of least resistance of an outflow, this may 
provide a further source of scatter in the metallicity of the IGM observed by
 \citet{2003ApJ...596..768S,2004ApJ...606...92S,psa06} and as yet unexplained
 in simulations. In a subsequent  paper in this series \citep{pgm07} 
 we will investigate these issues by performing
a direct comparison between these observations and synthetic QSO absorption
spectra produced using our analytic description of anisotropic outflows.

The strongest and weakest points of our method are the high dynamical range 
and the lack of gravitational dynamics, respectively. On the one hand,
the ratio of the largest to smallest mass scale we consider is
$M_{10}/M_1\approx50,000$. It would be very challenging for a numerical
simulation to achieve such a large dynamical range in mass
(that is, simulating  $L_*$ galaxies {\it and\/} dwarf galaxies
together), while having
sufficient resolution to properly simulate the outflows originating
from the smallest galaxies. On the other hand, the treatment of large-scale
structure and galaxy formation in the Monte Carlo method is quite simplistic.
We combine a Gaussian random density field with a spherical collapse
model for galaxy formation. In this approach, galaxies form at the
comoving locations of density peaks and remain at these locations
afterward. Even though we have a prescription for destruction of galaxies
by mergers, the actual clustering of galaxies is not taken into account.
If galaxies were allowed to cluster, collision and stripping by outflows
would be more frequent, and also it would become more difficult to 
enrich low-density regions with metals. This lack of a correct description
of dynamics means that the description of clustering of the IGM is also
limited.

The main limitation of this work is not the outflow model,
but rather the Monte Carlo model used for describing galaxy formation.
This Monte Carlo model has been used as a test bed for the outflow model in 
order to perform an investigation of its importance and potential influence.
In two forthcoming papers \citep{mgp07,pgm07}, we will replace this Monte Carlo
model by a more realistic numerical simulation of galaxy formation in a 
cosmological volume.

\section{SUMMARY AND CONCLUSION}

We have designed an analytical model for anisotropic galactic outflows
based on the hypothesis that such outflows are bipolar and follow
the path of least resistance through the environment of their source.
In this analytical model we vary one parameter: the opening angle, $\alpha$.
We combined this model with an analytical Monte Carlo method for
simulating galaxy formation, galaxy mergers, and supernova feedback.
With this combined algorithm, we study the evolution of the galaxies and
the IGM inside a comoving cosmological volume of size
$(12h^{-1}{\rm Mpc})^3$, from redshifts, $z=24-2$, in a
$\Lambda$CDM model. Our main results are the following:

\begin{itemize}

\item
Galaxy formation starts at redshift, $z\sim18$. Since we neglect the
formation and evolutionary times of massive stars, each newly formed galaxy
immediately produces an outflow that lasts for a time,
$t_{\rm burst}\sim50{\rm Myr}$. Such outflows can travel hundreds of 
kiloparsecs, and eventually collide with other objects. We neglect the effect
of a collision with a well-formed galaxy (the cross-section is
too small). When an outflow collides with a peak still in the
process of collapsing, removal of the gas by ram pressure, preventing
the formation of the galaxy, occurs about half of the time. When stripping
does not occur, the protogalaxy is enriched in metals. This process
occurs for all opening angles and the proportion  stripped or 
metal-enriched is essentially independent of opening angle.

\item
When metal-enrichment of a peak occurs and this peak collapses to form
 a halo, the cooling time of that halo is reduced,
and the galaxy forms earlier. However, this effect is rather small. In
particular, we did not find that metal-enrichment could
``bring to life'' low-mass protogalaxies whose cooling
time exceeds the age of the universe.

\item
Anisotropic outflows channel matter preferentially into low-density regions,
away from the cosmological structures (filaments or pancakes) in which the
galaxies producing the outflows reside. Consequently, the number of
halos encountered by outflows decreases with decreasing opening angle. This reduction
in the number of hits results in a larger number of galaxies forming, since
fewer halos are stripped by the ram pressure of outflows.

\item
The volume filling factor of galactic outflows (that is, the volume fraction of the IGM
occupied by outflows) holds constant and then decreases with opening angle.
For angles $\alpha=180^\circ-110^\circ$ the constant filling factor is a result of the 
balance between an increase due to larger numbers of galaxies and a decrease 
due to a fall in volume per outflow. At smaller angles, the
volume of individual outflows drops significantly with $\alpha$, and the
total filling factor decreases since this term wins out.

\item
The decrease in filling factor with decreasing angle is not sufficient
to explain the decrease in number of hits. The ratio 
(number of hits)/(filling factor) decreases with decreasing angles down to
$\alpha\sim50^\circ$. This indicates that at these angles, the outflows
are efficient at avoiding collisions with halos and
channel matter preferentially into low-density region. Hence, if several
halos reside in a common cosmological structure, an outflow produced by one
of them will tend to avoid encountering the others. For angles $\alpha<50^\circ$,
we observe the opposite trend: outflows become more efficient in finding halos
and hitting them. These narrow outflows can travel across cosmological voids
and hit halos located in unrelated structures, like the next filament or 
pancake. This effect is a continuation of a process begun at $\alpha\sim 100^\circ$
where the mean distance travelled by outflows when they strip collapsing 
peaks of their baryons begin to increase as the first neighboring structures
are hit.

\item
The enrichment of the IGM with metals favors high-density systems since the
sources of outflows are located in high-density regions. However, as the
opening angle decreases, there is a dramatic reduction of the enrichment of 
such systems, combined with a dramatic increase in enrichment of low-density
systems. Anisotropic outflows enrich around 10\% larger a volume of the 
underdense Universe (and $40\%$ more of the Universe below $\rho/{\bar \rho}=0.1$) 
 than isotropic outflows. 

\end{itemize}

This collection of results is a mere consequence of the fact
that outflows follow the path of least resistance. This is an assumption
in our model and is motivated by observations as well as high-resolution
simulations.

\acknowledgments

This work benefited from stimulating discussions with A.~Ferrara,
E.~Scannapieco, J.~Silk, M.~Tegmark, R.~Thacker, and S.~D.~White.
All calculations were performed at the {\sl Laboratoire
d'astrophysique num\'erique}, Universit\'e Laval. Figure~\ref{pancake}
was produced at the {\sl Center for Computer Visualization}, University of
Texas, by Marcelo Alvarez.
We thank the Canada Research Chair program and NSERC for support.
CG is also supported by a Hubert~Reeves fellowship.

\appendix

\section{NUMERICAL SOLUTION FOR THE OUTFLOW}

The equations governing the evolution of the outflow are
\begin{eqnarray}
\label{rdotdot2}
\ddot R&=&{8\pi (p-p_{\rm ext})G\over\Omega_bH^2R}-{3\over R}(\dot R-HR)^2
-{\Omega H^2R\over2}-{GM\over R^2}\,,\\
\label{pdot2}
\dot p&=&{L(t)\over2\pi R^3[1-\cos(\alpha/2)]}-{5\dot Rp\over R}\,.
\end{eqnarray}
We solve these equations numerically, starting with the initial
condition $R=0$ at $t=0$, which we have chosen for conceptual simplicity. We might 
have chosen our initial radius at $t=0$ to be any value below the physical size of 
the star-forming region of the source galaxy; however, the results are not 
substantially changed by such an adjustment. Also it is more plausible 
to expect that the starting point of the coherent outflows is closer to $R=0$ than the 
radius of the star-forming region, if we assume that the star formation density will 
be centrally peaked.

Most terms in equations (A1) and (A2)
diverge at $t=0$, making it impossible to obtain a numerical solution in 
this form. We must
first perform a change of variables that will eliminate the divergences
in these equations while retaining all the terms. To find the appropriate
change of variables, we investigate the early-time behavior of the solution,
by taking the limit $t\rightarrow0$. In this limit, we have
$R\rightarrow0$,
$H\rightarrow H_i$, $\Omega_b\rightarrow\Omega_{b,i}$,
$\Omega\rightarrow\Omega_i$, and $L\rightarrow L_i$.
The gravity term, Hubble 
terms, and external pressure term are negligible (in the sense that they
diverge slower than the other terms), and 
equation~(\ref{rdotdot2}) reduces to
\begin{equation}
\label{rdotdot0}
\ddot R={8\pi pG\over\Omega_{b,i}H_i^2R}-{3\dot R^2\over R}\,,\qquad
t\rightarrow0\,.
\end{equation}

\noindent
We can easily show that the solutions of the system of equations 
(\ref{pdot2})--(\ref{rdotdot0}) are power laws,
\begin{eqnarray}
R&=&Ct^{3/5}\,,\\
p&=&{21\Omega_{b,i}H_i^2C^2\over200\pi G}t^{-4/5}\,,
\end{eqnarray}

\noindent where
\begin{equation}
\label{cc}
C=\left\lbrace{500GL_i\over231\Omega_{b,i}
H_i^2[1-\cos(\alpha/2)]}\right\rbrace^{1/5}\, and \quad
t\rightarrow0\,.
\end{equation}

Using equations (A4) and (A5), we can find the proper change of variables. 
We introduce the following transformations (which are free from 
the above approximations),
\begin{eqnarray}
\label{sdef}
S&=&Rt^{2/5}\,,\\
\label{qdef}
q&=&pt^{9/5}\,,\\
\label{udef}
U&=&\dot St\,.
\end{eqnarray}

\noindent In the limit $t\rightarrow0$, the three functions
$S$, $q$, and $U$ vary linearly with $t$. We stress that this leaves the
 behavior of the outflow expansion unchanged and 
will only lead to a well-behaved system of equations at the start of the first time
step. We now eliminate the functions $R$ 
and $p$ in our original equations (A1) and (A2) using (A7)--(A9), and get
\begin{eqnarray}
\label{qdot2}
\dot q&=&{19q\over5t}+{Lt^3\over2\pi[1-\cos(\alpha/2)]S^3}-{5qU\over St}\,,\\
\label{sdot2}
\dot S&=&{U\over t}\,,\\
\label{udot2}
\dot U&=&{9U\over5t}-{14S\over25t}+{8\pi(q-q_{\rm ext})G\over\Omega_bH^2S}
-{3t\over S}\left({U\over t}-{2S\over5t}-HS\right)^2-{\Omega H^2St\over2}
\nonumber\\&&
-{GMt^{11/5}\over S^2}\,.
\end{eqnarray}

\noindent These equations are completely equivalent to our original equations
(A1) and (A2), but all divergences at $t=0$ have been eliminated.
These equations can therefore
be integrated numerically using a standard Runge-Kutta algorithm, with the
initial conditions $S=q=U=0$ at $t=0$. However,
before doing so, it is preferable
to rewrite the equations in dimensionless form. We define
\begin{eqnarray}
\label{taudef}
\tau&\equiv&H_0t\,,\\
\label{stdef}
\tilde S&\equiv&H_0S/C\,,\\
\tilde U&\equiv&H_0U/C\,,\\
\label{qtdef}
\tilde q&\equiv&Gq/H_0C^2\,,\\
f_H&\equiv&H/H_0\,,\\
f_L&\equiv&L/L_i\,,\\
{\cal M}&\equiv&GM/C^3H_0^{1/5}\,.
\end{eqnarray}

\noindent Equations~(\ref{qdot2})--(\ref{udot2}) 
reduce to their dimensionless form
\begin{eqnarray}
\label{qtdot}
{d\tilde q\over d\tau}&=&{19\tilde q\over5\tau}
+{231\Omega_{b,i}f_Lf_{H,i}^2\tau^3\over1000\pi \tilde S^3}
-{5\tilde q\tilde U\over\tilde S\tau}\,,\\
{d\tilde S\over d\tau}&=&{\tilde U\over\tau}\,,\\
{d\tilde U\over d\tau}&=&{9\tilde U\over5\tau}-{14\tilde S\over25\tau}
+{8\pi(\tilde q-\tilde q_{\rm ext})\over\Omega_bf_H^2\tilde S}
-{3\tau\over\tilde S}
\left({\tilde U\over\tau}-{2\tilde S\over5\tau}-f_H\tilde S\right)^2
-{\Omega f_H^2\tilde S\tau\over2}
\nonumber\\&&
-{{\cal M}\tau^{11/5}\over\tilde S^2}\,,
\end{eqnarray}

\noindent where $f_{H,i}=H_i/H_0$. Interestingly, 
the change of variable eliminates the explicit dependence upon the
opening angle in equation~(\ref{qtdot}).

The quantity $f_L$ appearing in equation~(\ref{qtdot}) depends on
the luminosity $L_{\rm comp}$, which is given by equation~(\ref{lcomp2}).
We eliminate $p$ and $R$ in equation~(\ref{lcomp2}), using 
equations~(\ref{sdef}), (\ref{qdef}), (\ref{taudef}), (\ref{stdef}), 
and~(\ref{qtdef}), then eliminate $C$ using equation~(\ref{cc}). We get
\begin{equation}
{L_{\rm comp}\over L_i}={200\pi^3\over2079}
{\sigma_t\hbar\over m_eH_0\Omega_{b,i}f_{H,i}^2}
\left({kT_{\gamma0}\over\hbar c}\right)^4
(1+z)^4{\tilde q\tilde S^3\over\tau^3}\,.
\end{equation}

The initial conditions are $\tilde S=\tilde q=\tilde U=0$ 
at $\tau=0$.
In the limit $\tau\rightarrow0$, where many terms take the form
$0/0$, the derivatives reduce to 
$d\tilde S/d\tau=d\tilde U/d\tau=1$, and 
$d\tilde q/d\tau=21\Omega_{b,i}f_{H,i}^2/200\pi$. We solve these equations 
numerically, using a fourth-order Runge-Kutta algorithm.

%

\end{document}